\documentclass[11pt,english]{article}
\usepackage{lmodern}

\usepackage[T1]{fontenc}
\usepackage[latin9]{inputenc}
\usepackage{geometry}
\usepackage{array}
\usepackage{float}
\usepackage{amsmath}
\usepackage{amsthm}
\usepackage{amssymb}
\usepackage{graphicx}
\usepackage{setspace}
\usepackage[authoryear]{natbib}
\usepackage{hyperref}
\usepackage{tabularx}
\usepackage{siunitx}

\makeatletter

\providecommand{\tabularnewline}{\\}

\setcitestyle{round}
\usepackage{lscape}
\usepackage{longtable}

\usepackage{graphicx}
\usepackage{tabularx}
\newcolumntype{P}[1]{>{\centering\arraybackslash}p{#1}}
\newcolumntype{Y}{>{\centering\arraybackslash}X}

\usepackage{array,booktabs,ragged2e}
\newcolumntype{C}[1]{>{\centering\arraybackslash}p{#1}}
\newcolumntype{J}[1]{>{\justify\arraybackslash}p{#1}}
\newcolumntype{R}[1]{>{\RaggedLeft\arraybackslash}p{#1}}
\newcolumntype{Q}[1]{>{\columncolor{Gray}\RaggedLeft\arraybackslash}p{#1}}
\newcolumntype{L}[1]{>{\RaggedRight\arraybackslash}p{#1}}
\newcolumntype{G}{@{\extracolsep{0.5cm}}l@{\extracolsep{0pt}}}%
\usepackage{multirow}

\AtBeginDocument{

}

\makeatother

\usepackage{babel}
\begin{document}
\title{Periodicity in Cryptocurrency Volatility and Liquidity\thanks{We are grateful to Tim Bollerslev, Chris Cashwell, and Mark Ettinger for many useful and constructive comments and to Emily Dyckman for proofreading the article.} \thanks{Disclosure: The first and second authors have served as compensated consultants for Volatility Group Inc.}\vspace{1cm}}
\author{
    \textbf{Peter Reinhard Hansen}$^{a,b}\thanks{Corresponding author. Address: University of North Carolina, Department of Economics, 107
Gardner Hall Chapel Hill, NC 27599-3305}\quad$ 
    \textbf{Chan Kim}$^{a}\quad$
    \textbf{Wade Kimbrough}$^{c}$
    \bigskip{}\bigskip{}\\
{$^{a}$}\emph{University of North Carolina, Chapel Hill\medskip{}}\\
{\normalsize{}$^{b}$}\emph{Copenhagen Business School\medskip{}
}\\
{\normalsize{}$^{c}$}\emph{Volatility Group, Inc.\medskip{}
}}
\date{\emph{\today}}
\maketitle
\begin{abstract}
\noindent{}We study recurrent patterns in volatility and volume for major cryptocurrencies, Bitcoin and Ether, using data from two centralized
exchanges (Coinbase Pro and Binance) and a decentralized exchange (Uniswap V2).
We find systematic patterns in both volatility and volume across day-of-the-week, hour-of-the-day, and within the hour. These patterns have grown stronger over the years and are presumably related to algorithmic trading and funding times in futures markets. We also document that price formation mainly takes place on the centralized exchanges while price adjustments on the decentralized exchanges can be sluggish.

\bigskip{}
\end{abstract}

\noindent \textit{\small{}JEL Classification:}{\small{} C58, G12 \newpage}{\small\par}

\section{Introduction}

We study empirical patterns in volatility in cryptocurrency markets.
Intraday patterns in volatility are well-documented in stock prices
and in foreign exchange rates, see \citet{Andersen1997}, \citet{Andersen1998b},  \citet{AndersenThyrsgaardTodorov:2019}, and references therein. The average time-of-the-day volatility for US stock returns is typically found to have a U-shape over the period from 9:30 AM to 4:00 PM.
It is also well documented that volatility tends to be
higher in the hours during which main exchanges are open. For instance,
\citet{HansenHuangShek:2012} found that roughly 75\% of daily volatility
in the S\&P 500 index occurs during the 6.5 hours where US stocks
are actively traded on major exchanges. 

Cryptocurrencies differ from most other markets by being open for
trading 24 hours a day, seven days a week. Trading activity is continuously high, and therefore, high-frequency data is available at any time of the day for all major cryptocurrencies.
This availability of data facilitates studies of periodicity in volatility and market
liquidity measures beyond what is possible in other markets. 

In this paper, we investigate periodicity in volatility and liquidity in
two leading cryptocurrencies, Bitcoin and Ether, using data from
three exchanges, Binance, Coinbase Pro, and Uniswap V2. Periodicity refers to a recurrent pattern within some period, such as a week, a day, or an hour.\footnote{Parts of the literature refer to such patterns as seasonality even though the patterns are unrelated to seasons, and the periods are shorter than a year.} 
We find strong evidence of periodicity in volatility and liquidity for day-of-the-week, hour-of-the-day, as well as within the hour. The first two findings are not new, but the
distinct patterns we document within the hour are new. We conjecture
that many of these patterns are driven by algorithmic trading while some are likely related to funding rates for perpetual futures.

The recurrent patterns have important implications for volatility measurement, modeling and forecasting. First, in measuring volatility it is valuable to know if a sudden change in volatility merely follows the expected pattern or if it reflects an unanticipated change. This importance is analogous to the way macroeconomic time series are adjusted for seasonality. For example, to understand a measurement of volatility prior to the release of an unemployment report, one would need to know the difference between the current measurement and the pattern of high volatility that has been established during the lead up to previous reports being released.  Similarly, knowledge about the typical variation in trading volume makes it possible to discriminate between usual variation and unanticipated changes. Second, in time series forecasting we can account for periodicity, analogous to the way seasonality is built into models. In our empirical analysis we find that GARCH models benefit substantially by accounting for periodicity. The estimated models with periodicity have substantially better in-sample and out-of-sample performance than standard models without periodicity. Third, some of the recurrent patterns we identify may also have important implications for some realized measures of volatility as we discuss in the conclusion.

Recurrent patterns in cryptocurrency markets have previously been documented
in \citet{Dimpfl2017} who found that trading volume was closely related
to the trading times of the US and European stock markets. Similarly,
\citet{DyhrbergFoleySvec2018} found that the highest trading activity,
highest volatility, and lowest spread in three US based crypto-exchanges
coincide with US market trading hours. \citet{CataniaSandholdt2019}
documented that the intraday volatility and volume patterns were different for a Europe-based exchange (Bitstamp) and a US-based exchange (Coinbase), and that the patterns followed the time-of-the-day in their respective locations.
In terms of the intra-week patterns, both exchanges show increasing
volatility during weekdays which then decreases on weekends. \citet{AletiMizrach2021}
conclude that crypto-liquidity generally peaks when the US and London stock markets
are both open whereas the periods of lower crypto-volume and wider spreads coincide with the hours where the CME Bitcoin futures market is closed. \citet{WangLiuHsu:2020} relate patterns in volatility and volume to stock market openings in Asia, Europe, and US. 
Patterns in US stock markets have been related to algorithmic trading activities at an ultra high time-scale. For instance, \citet{HasbrouckSaar:2013} documented patterns within the second, with higher trading activity during the early milliseconds of the second with a peak around the 30th millisecond and another smaller spike around the 150th millisecond. These patterns were attributed to algorithmic trading, latency, and geographical clustering of algorithmic traders, as further documented by \citet{DobrevSchaumburg:WP}.

Our investigation of periodicity of volatility and liquidity of cryptocurrencies is laid out in seven sections. In Section 2, we provide a brief introduction to cryptocurrencies and cryptocurrency markets, and we highlight the key characteristics of the decentralized exchange used in our analysis. Decentralized exchanges are a relatively new concept. While they have many potential applications, their key use is the trading of cryptocurrencies.\footnote{At the time of writing we are only aware of a handful of betting platforms that operate as decentralized exchanges.} We describe our data in Section 3 along with descriptive statistics and some preliminary results. In Section 4, we compute cross correlations between exchanges and currencies that strongly indicate that price formation takes place on centralized exchanges. We establish the "seasonal'' patterns in volatility, volume, and illiquidity in Section 5. In Section 6, we show that accounting for periodicity in daily returns improves the empirical fit of volatility modeling and forecasting, and our concluding remarks are presented in Section 7. Two Appendices include additional empirical results and theoretical justification of the confidence intervals we present.

\section{Cryptocurrency and Exchanges}

Cryptocurrency is digital money, sometimes considered more generally as a digital asset.  The core idea behind cryptocurrency is that the creation, distribution, and transfer of a cryptocurrency is \emph{decentralized}, i.e. not controlled by a centralized authority such as a central bank.  This feat is accomplished using an innovative database called a \emph{blockchain} and transactions secured by computationally strong public-key cryptography.  See \cite{NarayananEtAl2016} for an introduction and overview of cryptocurrency technology.

Though there are thousands of cryptocurrencies, this paper focuses on five: Bitcoin (symbol BTC), Ether (symbol ETH), Wrapped Ether (symbol WETH), USD Coin (symbol USDC), and USD Tether (symbol USDT). Using these five cryptocurrencies allows us to investigate the periodicity of volatility of BTC and ETH from realized prices (in U.S. Dollars) across our chosen centralized and decentralized exchanges. We selected these tokens because BTC and ETH are the most traded, have the largest market caps, and are the best known cryptocurrencies. We paired each of these with a USD counterpart to investigate their volatility in relationship to the US Dollar and measure their trading volume in USD.

Bitcoin was the first and, at the time of writing, the largest cryptocurrency by market capitalization, accounting for 45\% of the cryptocurrency market.  The utility of Bitcoin is primarily as a form of digital money (though ``money'' in this context is not without controversy, see \cite{Taleb2021}) due to the fact that the underlying blockchain database is in essence only a financial ledger. Albeit, this is a slight oversimplification, see \cite{NarayananEtAl2016}.

At the time of writing, Ether is the second largest by market capitalization making up 18\% of the market.  Ether is the currency underlying Ethereum, a decentralized virtual computer. On Ethereum, as with Bitcoin, transactions are added to the blockchain when confirmed through a consensus mechanism. \emph{Miners} compete to solve a difficult algorithm to add the next \emph{block} to the current chain, where the block consists of a batch of transactions. On average, a block is added every 13 seconds. In order to get a transaction into a block, a user must bid on \emph{gas} (a fee) denominated in Ether.\footnote{Ethereum executes programs, called \emph{smart contracts}, consuming \emph{gas}. Gas is the fee, denominated in Ether, which is required to transact on Ethereum.} Transactions are typically ordered in the block by how much a user pays in gas (e.g. the more someone pays in gas the higher it is in the order). The Ethereum blockchain does not just store financial balances, but rather the state of the Ethereum Virtual Machine (EVM), a virtual computer which ``runs'' on physical computers all over the world. For more detail, refer to  \cite{ethorg}.

WETH is a proxy cryptocurrency representing Ether in the form of an ERC-20 token,\footnote{For detailed descriptions of WETH and ERC-20, we refer to \url{https://weth.io/} and \url{https://eips.ethereum.org/EIPS/eip-20}, respectively.} the Ethereum standard for fungible tokens. WETH is created by swapping ETH into a smart contract at a 1:1 ratio.  Because WETH trades on Uniswap V2 and ETH does not, WETH is the source of decentralized prices in this paper.

Cryptocurrencies pegged to fiat currencies are typically referred to as \emph{stablecoins}. Stablecoins may be either centralized or decentralized. Centralized stablecoins are issued by a regulated entity which backs the stablecoin with fiat currency either partially or fully. Decentralized stable coins are issued algorithmically on blockchain and are collateralized with a variety of assets. 

Two centralized stablecoins representing the US Dollar are USDC and USDT.  USDC is a stablecoin issued by Circle Internet Financial LLC. It is regulated and backed 1:1 with assets that total in USD value the amount of USDC in circulation. USDC is a trusted, centralized stable coin because its reserves are independently audited. USDT is a digital asset issued by Tether Operations Limited. There is controversy surrounding the adequacy of USDT reserves because, at the time of writing, USDT reserves have not been independently audited.  However, independent accountants have examined USDT reserves.

As is the case for stock and foreign exchange markets, trading of cryptocurrencies is fragmented across multiple exchanges. In this paper, we use data from three exchanges: Binance, Coinbase Pro, and Uniswap V2. Binance and Coinbase Pro are centralized exchanges (CEX), while Uniswap V2 is a decentralized exchange (DEX). We will describe each in turn while putting an emphasis on the decentralized exchange because it uses mechanics that are quite different from traditional financial exchanges.

Centralized exchanges are platforms owned and operated by a legally regulated entity. They facilitate the exchange of cryptocurrencies for fiat currencies or other cryptocurrencies. They act as financial custodians where users deposit their assets. Mechanically, they are very similar to traditional financial exchanges. Bid and ask prices are matched in an order book.  Arbitrageurs and market makers play a central role in keeping price parity between exchanges. Centralized exchanges fall under the purview of financial regulators from where they are incorporated and from the territories in which their customers reside. Therefore, centralized exchanges often ban users from regions where they cannot, or are unwilling to, comply to certain regulations. The geographic location of traders may affect how volume and volatility is distributed over the day, which is one of several patterns we study in this paper.

Binance launched in June 2017 and is the largest centralized exchange in the world by daily volume at the time of this writing.\footnote{\$55 Billion USD on May 31, 2021 according to crypto-data provider Nomics ( \url{https://nomics.com/})} The currency pairs  BTC/USDT and ETH/USDT had the highest trading volume on Binance during the first half of 2021. Web-traffic to Binance is widely distributed across the world and, as of May 2021, Turkey had the largest share (7.6\%) followed by the UK (4.66\%).\footnote{\url{https://www.similarweb.com/website/binance.com/}} For regulatory reasons, Binance has not served US residents since September 2019 and has not served UK residents since June of 2021. A separate exchange, Binance US, serves US residents.
 
Coinbase Pro is ranked 15th by volume (\$3.62 Billion USD according to Nomics on May 31, 2021) among centralized exchanges and is the largest cryptocurrency exchange in the US. Their BTC/USD and ETH/USD markets both have the largest trading volume across all BTC and ETH markets with U.S. Dollars. Unlike Binance, the web-traffic to Coinbase Pro is very concentrated in the US (56.64\%) followed by the UK (7.33\%), as of May, 2021.\footnote{\url{https://www.similarweb.com/website/pro.coinbase.com/}}  

In contrast to centralized exchanges, decentralized exchanges (DEXs) are not owned or operated by any entity.  Decentralized exchanges are sets of smart contracts, i.e. computer code, that  implement financial protocols.  Users can create markets or trade cryptocurrencies using these smart contracts.  Most of the protocols for DEXs are known as Automated Market Makers (AMMs). There are several different AMM protocols on Ethereum and they each employ different mechanics. For example, one might use a constant sum formula to determine price, while another might use a constant product formula to determine price.  For a more thorough overview of AMM protocols, see \cite{XuEtAl2021}.

Uniswap V2 is a general protocol implemented in smart contracts to facilitate trading one ERC-20 token for another (e.g. USDC/WETH). Uniswap employs a constant product formula to govern the dynamics of liquidity provision and currency trading.  A constant product formula has the form \(K = X\times Y\), where \(X\) and \(Y\) are the reserves of each cryptocurrency and \(K\) is a constant which remains invariant during operation of the AMM (with the exception of accumulated trading fees).  Any user can create a market by initially depositing both cryptocurrencies into a pool that is governed by the Uniswap smart contract. The ratio of the assets determines the spot price. Because of this, these exchanges are more commonly referred to as \emph{pools}.  The amount of each token in the pool is referred to as  \emph{liquidity}. Any user can add liquidity to a pool by depositing the two cryptocurrencies according to the current reserve ratio. For example, if a pool contains 20,000 USDC and 10 ETH, liquidity providers add additional liquidity at the same ratio (2,000 USDC and 1 ETH or 200 USDC and 0.1 ETH, etc.). Liquidity providers can add and remove liquidity without paying a transaction fee but must pay the cost of gas. A 0.3\% fee is charged for swapping one cryptocurrency for another. This fee is added to the currency reserves in the pool, thus providing an incentive to liquidity providers.

On Uniswap V2, anyone can interact with a smart contract to swap one cryptocurrency for another. The net exchange rate is determined by the constant product formula defining that \(x_\mathrm{in}\) of one cryptocurrency can be swapped for \(y_\mathrm{out}\) of another currency according to the formula:
\[y_\mathrm{out} = Y - \frac{K}{X + (1-0.003)x_\mathrm{in}}=\frac{1}{\frac{1}{0.997}+\frac{x_\mathrm{in}}{X}}\times \frac{Y}{X}\times x_\mathrm{in},\]and it follows that the average price for this transaction is 
\[
\frac{x_\mathrm{in}}{y_\mathrm{out}} = 
\left(\frac{1}{0.997}+\frac{x_\mathrm{in}}{X}\right)\times \frac{X}{Y} \label{eq:TransactionPriceDecentralized}
\]
This clarifies the transaction cost because, in the absence of transaction costs, the price of one currency denominated in the other currency would simply be the ratio $X/Y$. Unlike an order book, the spot price, $X/Y$, plus the fee only holds "infinitesimally" on Uniswap V2 because any finite trade involves a continuous movement of the reserves along the constant function curve (from 0 to $x_\mathrm{in}$). Therefore, the greater the quantity of currency exchanged, the more the realized price deviates from the initial spot price before the trade.  This deviation in price is called \emph{slippage}, and is directly related to the amount exchanged and the total liquidity in the pool, as captured by the term \(\tfrac{x_\mathrm{in}}{X}\). The concept of slippage is important in our investigations because it affects the dynamics of trading volume and volatility on Uniswap V2 with respect to Binance and Coinbase Pro.

\section{Data and Preliminary Results}

We include data for five markets: BTC/USD (Coinbase Pro), ETH/USD (Coinbase Pro), BTC/USDT (Binance), ETH/USDT(Binance), and WETH/USDC (Uniswap V2). 

We downloaded 1-minute OHLCV data\footnote{OHLCV = Open, High, Low, Close prices and traded Volume.} and trade data 
for the two centralized exchanges, Binance and Coinbase Pro, through \texttt{nomics.com} using their API (for a fee).\footnote{We also obtained the same OHLCV data using the Python package CCXT by \cite{CCXT} and verified that the data is identical.}
OHLCVs per Ethereum block for Uniswap
V2 are collected through a freely available API at \url{graphql.bitquery.io}.
The block-by-block OHLC prices of Uniswap V2 are different from those
of the orderbooks. Each block takes 13 seconds on average with less than 1\% of the blocks spanning more than 1 minute. Each block consists of many ordered transactions, and we define our price to be the ratio of the final balances at the end of the block of the two cryptocurrencies within the liquidity pool. This choice circumvents the impact that fees and slippage have on transaction prices, see Section \ref{eq:TransactionPriceDecentralized}, and mitigates the impact that manipulative attacks can have on our price series.
It is possible to carry out several types of attacks to manipulate prices on decentralized exchanges, especially within the block. Pools with high slippage are especially vulnerable to price manipulation. 
Specifically, we calculate
1 minute returns of the WETH/USDC market on Uniswap V2 by the log difference of the last marginal price of the last block for each minute. We do not use Uniswap V2 prices for BTC because the volume is insufficient for a detailed analysis based on high frequency data.

We present detailed results for the sample period from October 1, 2020 at 00:00 UTC to June 1, 2021 at 00:00 UTC, a period where high-frequency data is continuously available for all exchanges. We present additional results for the centralized exchanges based on an extended sample period from January 1st, 2019 at 00:00 UTC to June 1st, 2021 at 00:00 UTC.
Although the WETH/USDC market launched on Uniswap V2 in May of 2020 we focus on a shorter sample period for two reasons. First, there was insufficient trading activities during the initial months after Uniswap V2 was launched. Second, in late August of 2020 a protocol called Sushiswap launched. Sushiswap forked much of Uniswap's code and incentivized users to transfer their liquidity to the new protocol. This caused a large disruption in the depth of liquidity on Uniswap V2 with roughly 55\% of total liquidity being transferred to Sushiswap for several weeks. By October 1, 2020 liquidity of Uniswap V2 had returned to earlier levels.
We did not include data for the centralized exchanges before 2019 for two reasons. First, there were credibility issues associated with USDT that caused USDT to deviate from the value of the US Dollar. Second, Binance had a small market share before 2019.

\subsection{Realized Volatility Measures}

We compute the daily realized variance with five-minute returns,
\begin{equation}
\mathrm{RV}_{t}=\sum_{j=1}^{288}(Y_{t,5j}-Y_{t,5j-5})^{2},\label{eq:RV}
\end{equation}
where $Y_{t,i}$ is the logarithmic price of a cryptocurrency at the $i$-th
minute on day $t$, $i=0,\ldots,1440$.  The annualized volatility time series are defined by: 
\[
\sqrt{365\times\mathrm{RV}_{t}}\times100\%.
\]
To calculate the realized variance, we use five-minute returns
rather than one-minute returns because sparse sampling offers
greater robustness to market microstructure noise. Market microstructure noise will reveal itself as autocorrelation in high-frequency returns, which we estimate below.

Table \ref{tab:Summary} presents summary statistics about daily annualized volatility based on 5-min returns and daily volumes. Binance and Coinbase Pro show similar price movement for both BTC and ETH, as can be seen from the sample correlation matrix below. For this reason, their average annualized volatilities are nearly identical for both BTC ($\sim$88\%) and ETH ($\sim$111\%). The corresponding volatility measure for ETH using Uniswap V2 data is substantially smaller at just 90.66\%. This is caused by a higher level of noise in the prices obtained from Uniswap V2 as we discuss below. The average daily volume is also substantially smaller for Uniswap V2 than the two centralized exchanges. This bolsters the argument that the realized variances from the centralized exchanges are more reliable than realized variances from Uniswap V2 data.
\begin{table}[H]
\centering{}\begin{small}
\begin{tabular*}{1\textwidth}{@{\extracolsep{\fill}}>{\raggedright}p{0.26\textwidth}rrrrrr} 
\hline
\\[-1pt]
\multicolumn{7}{c}{RV (Annualized)}
\\[4pt]
&\multicolumn{1}{c}{Average}&\multicolumn{1}{c}{min}&\multicolumn{1}{c}{q25}&\multicolumn{1}{c}{q50}&\multicolumn{1}{c}{q75}& \multicolumn{1}{c}{max}\\[3pt]
BTC/USDT (Binance)& \multicolumn{1}{c}{88.33} & \multicolumn{1}{c}{17.90} & \multicolumn{1}{c}{54.88} & \multicolumn{1}{c}{76.33} & \multicolumn{1}{c}{102.67} & \multicolumn{1}{c}{507.05}\\[2pt]
BTC/USD (Coinbase Pro)& \multicolumn{1}{c}{88.79} & \multicolumn{1}{c}{19.01} & \multicolumn{1}{c}{54.94} & \multicolumn{1}{c}{76.03} & \multicolumn{1}{c}{106.44} & \multicolumn{1}{c}{487.14}\\[2pt]
ETH/USDT (Binance)& \multicolumn{1}{c}{110.91}&\multicolumn{1}{c}{28.76}&\multicolumn{1}{c}{65.76}&\multicolumn{1}{c}{92.66}&\multicolumn{1}{c}{129.41}&\multicolumn{1}{c}{773.62}\\[2pt] 
ETH/USD (Coinbase Pro)& \multicolumn{1}{c}{111.21} &\multicolumn{1}{c}{27.82} & \multicolumn{1}{c}{65.67} & \multicolumn{1}{c}{92.36} & \multicolumn{1}{c}{131.32}   &\multicolumn{1}{c}{753.03}\\[2pt]
WETH/USDC (Uniswap V2)& \multicolumn{1}{c}{90.66} & \multicolumn{1}{c}{17.22} & \multicolumn{1}{c}{55.95} & \multicolumn{1}{c}{78.01} & \multicolumn{1}{c}{102.94} & \multicolumn{1}{c}{400.14}\\  
&  &  &  &  &  & \\
\multicolumn{7}{c}{Volume}
\\[4pt]  
& \multicolumn{1}{c}{Average} & \multicolumn{1}{c}{min} & \multicolumn{1}{c}{q25} & \multicolumn{1}{c}{q50} & \multicolumn{1}{c}{q75} & \multicolumn{1}{c}{max}\\[3pt]
BTC/USDT (Binance)& 3,034M & 235M & 1,513M & 2,951M & 4,133M & 13,482M\\[2pt]
BTC/USD (Coinbase Pro)& 829M & 34M & 333M & 726M & 1,182M & 4,177M\\[2pt]
ETH/USDT (Binance)& 1,735M & 105M & 539M & 1,386M & 2,278M & 11,645M\\[2pt]
ETH/USD (Coinbase Pro)& 539M & 14M & 102M & 361M & 712M & 4,510M\\[2pt]
WETH/USDC (Uniswap V2)& 205M & 18M & 51M & 127M & 188M & 3,337M\\  
&  &  &  &  &  & \\
\hline
\end{tabular*}
\end{small}\begin{small}\caption{Summary Statistics for the sample period October 1, 2020 to May 31,
2021.\label{tab:Summary}}
\end{small}
\end{table}

Next, we turn to the one-minute intraday returns, $y_{t,i}=Y_{t,i}-Y_{t,i-1}$,
for $i=1,\ldots,1440.$ The empirical correlation matrix for the five
time series of contemporaneous one-minute returns, for the sample period October 1, 2020 to May 31, 2021, is given by:

\begin{center}
  \begin{tabular}{@{} c  c c c c c >{\hspace{2mm}}l @{}} 
    \multirow{5}{*}{$\hat{C} = $}  & 
      \multicolumn{5}{@{}l}{
        \multirow{5}{*}{%
        $
        \begin{bmatrix}    
            1 & 0.823 & 0.969 & 0.802 & 0.322\\ 
            0.823 & 1 & 0.836& 0.968& 0.385\\ 
            0.969 & 0.803 & 1 & 0.800 & 0.328\\ 
            0.802 & 0.968 & 0.800 & 1 & 0.388\\ 
            0.322 & 0.385& 0.328& 0.388& 1     
        \end{bmatrix}
        $
      }
    } 
          & \scriptsize (BTC/USDT Binance) \\ 
& & & & & & \scriptsize (ETH/USDT Binance) \\
& & & & & & \scriptsize (BTC/USD Coinbase Pro) \\
& & & & & & \scriptsize (ETH/USD Coinbase Pro)\\
& & & & & & \scriptsize (WETH/USDC Uniswap V2)
  \end{tabular}\medskip{}
\par\end{center}
As expected, BTC/USD and BTC/USDT are highly correlated,
with a sample correlation of $0.969$, and similarly for ETH/USD and ETH/USDT
with a sample correlation of $0.968$. We also observe that Bitcoin and Ether returns are
very correlated, both within and across centralized exchanges, with correlations
varying between $0.802$ and $0.823$. The instantaneous correlations
between Ether returns on centralized exchanges and Ether returns on the
decentralized exchange are remarkably smaller at about 0.328. Perfect
competition would dictate the correlations to be near one for the
same cryptocurrency. However, sluggish price adjustments can cause
the correlations to be smaller. We do find evidence that prices on
the decentralized exchanges can be stale. Price changes on the centralized
exchanges are positively correlated with subsequent price changes on the
decentralized exchanges. This implies sluggish price revisions on Uniswap V2. Competitive market forces will only ensure that prices stay within a band, where the width of the band is defined by the cost of transacting. Price staleness is therefore a natural consequence of relatively high transaction costs. On Uniswap V2, the arbitrage must exceed the cost of the 
0.3\% transaction fee plus the cost of gas. The latter increases when many users compete to get their transaction into a block. The cost of gas is influenced by the amount of all Ethereum blockchain transactions not just Uniswap V2 transactions. Slippage determines if the cost of transacting will exceed the amount of arbitrage. 
We should acknowledge that some cross correlation can be explained by asynchronicity in the prices series. While the prices on the centralized exchanges come with microsecond precision, the duration between Ethereum blocks will induce some staleness in our Uniswap V2 prices. This can explain part of the cross-correlation between returns on the centralized exchanges and returns on Uniswap V2 the following couple of minutes, but the duration cannot explain the cross-correlations we observed 3 (and more) minutes later because block durations lasting more than 2 minutes are extremely rare, see Figure \ref{fig:BlockDurations} in the appendix. 
\begin{figure}[H]
\begin{centering}
\includegraphics[width=1\textwidth]{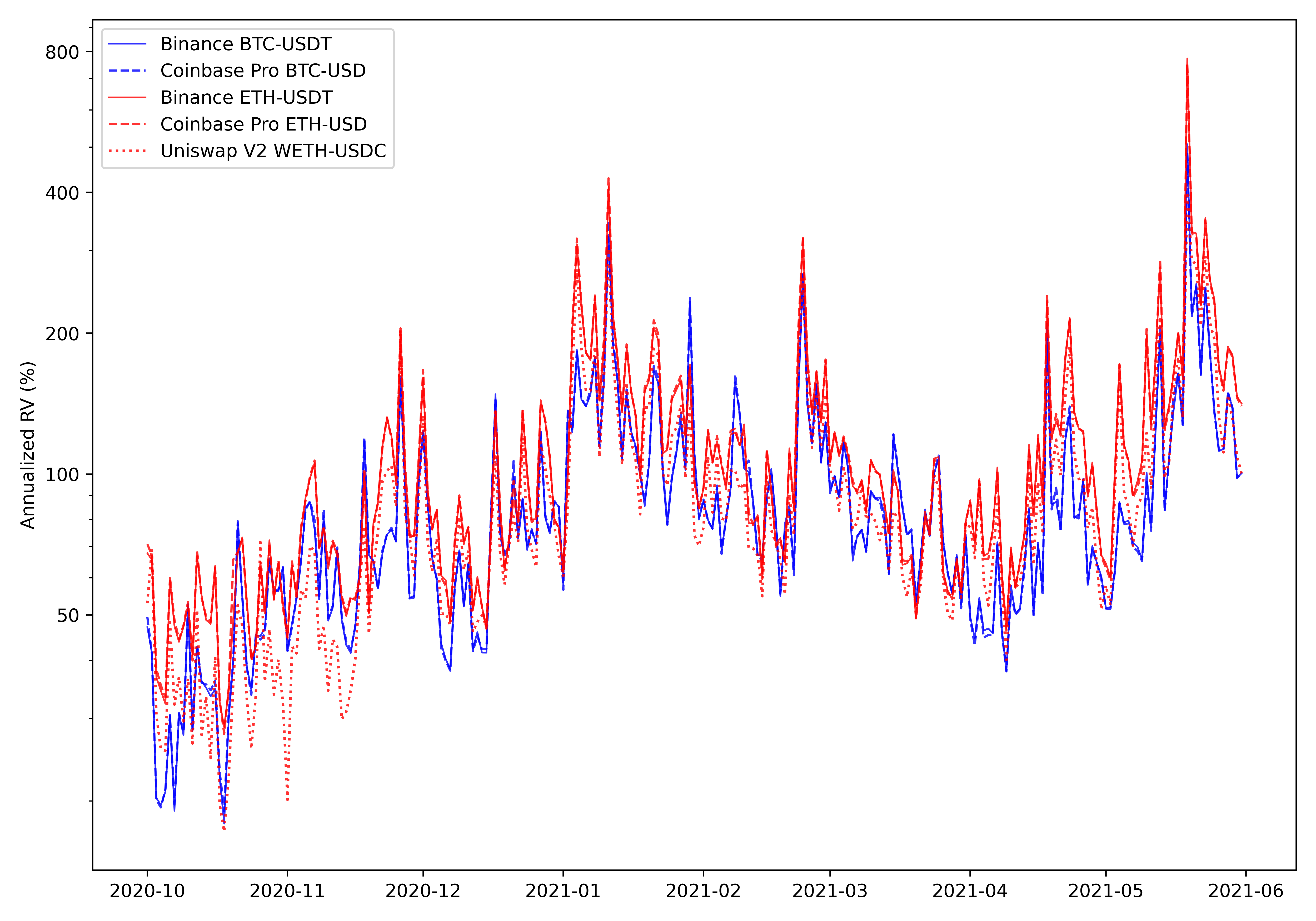}
\par\end{centering}
\begin{small}\caption{Daily realized variance based on five-minute returns for Bitcoin
(red) and Ether (blue). \label{fig:RealizedVariance5TimeSeries}}
\end{small}
\end{figure}

The time series for each of the five realized variances, in units of annualized volatility,  are presented in Figure \ref{fig:RealizedVariance5TimeSeries}.
The two volatility series for Bitcoin (solid blue line and dashed blue line) are virtually identical as one would expect if the underlying prices series were in agreement. The same is true for the volatility measures of Ether when computed from high-frequency prices from the centralized exchange (solid red line and dashed red line). The volatility series for Ether is quite different when computed from one-minute prices on the decentralized exchange, Uniswap V2. This is not surprising because the correlation between returns on the centralized exchanges and Uniswap V2 was only about 68\% whereas the correlation between returns on the two centralized exchanges was 99\%.
The volatility series for BTC and ETH vary similarly over time. This co-movement manifests a common risk component for the two cryptocurrencies as measured in relation to USD.

\section{Autocorrelations and Price Discovery}

The observed price process does not fully conform with properties
of a semi-martingale for a variety of reasons broadly known as market
microstructure effects. The literature on measuring volatility
from high-frequency data defines the discrepancy between the observed price
and the theoretical efficient price as market microstructure
noise, see \citet{HansenLundeRVmms}. The noise arises from bid-ask bounces, staleness, rounding errors, plain
data errors, and many other reasons. This type of noise reveals
itself as autocorrelations in high-frequency returns. Transaction costs can cause observed prices to be stale, and differences in staleness across exchanges will materialize as cross-correlation between intraday returns. 

\begin{figure}[H]
\begin{centering}
\includegraphics[width=1\textwidth]{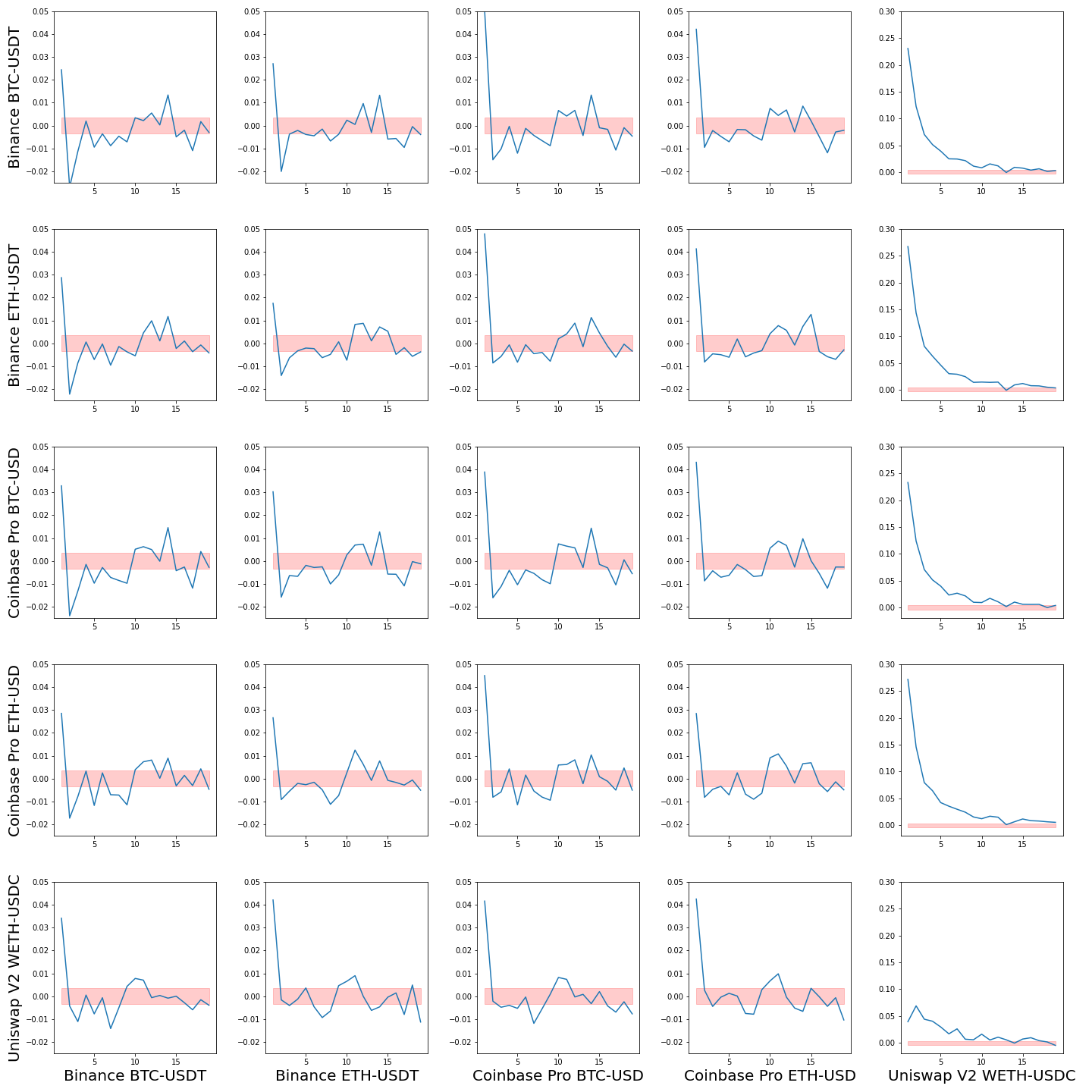}
\par\end{centering}
\caption{Autocorrelations and cross-correlations of one-minute returns for
the five currency pairs for $h=1,\ldots,20$. Shaded areas are 95\%
confidence bands. The first and second autocorrelations and cross-correlations tend to be significant for the centralized exchanges.
Price changes on Uniswap tend to lag price changes on the centralized exchanges
as evident from the many significant cross-correlations.\label{fig:Autocorrelations}}
\end{figure}

We present the autocorrelations and cross-correlations for the five
currency pairs in Figure \ref{fig:Autocorrelations} that are sample
estimates of $\rho_{k,l}(h)=\mathrm{corr}(y_{t,i}^{(k)},y_{t,i+h}^{(l)})$,
where $k$ (and $l$) refer to a particular currency/exchange, such as BTC/USDT on Binance.

The panels along the diagonal of Figure \ref{fig:Autocorrelations}
present the typical autocorrelation functions for a univariate time-series. These show price changes correlate with future price changes of the same price series.
For the centralized exchanges, the first two autocorrelations are statistically significant
in all cases, whereas the decentralized exchange has substantially
more positive autocorrelations. These empirical observations strongly suggest that liquidity
pool prices are sometimes sluggish and that price change can take several
minutes to be absorbed on the decentralized exchange, Uniswap V2.

The cross
correlations further support this view. For instance, the far right
panels, second and third from below, show that ETH intraday returns on
Binance and Coinbase Pro, $y_{t,i}^{(c)}$, are positively correlated
with subsequent ETH intraday returns on Uniswap V2, $y_{t,i+h}^{(d)}$, for
$h=1$ to about $h=12$. In fact, BTC price changes on the centralized exchanges predict future ETH price changes on Uniswap V2, as can be seen from the two top panels in the far right column of Figure \ref{fig:Autocorrelations}. The reason for this is the high correlation between BTC and ETH returns on the centralized exchanges. Some of the cross-correlations, especially for $h=1$, may be explained by the so-called Epp's effect which implies a downwards bias in contemporaneous correlations and an upwards bias in cross-correlations of the same magnitude. The staleness in prices on Uniswap V2 also induces this Epp's effect. Note the different $y$-scale that is used for cross-correlations in the far-right panels. These cross-correlations are much larger than those in the other panels, which reflects that future Uniswap V2 prices (ETH/USDC) are predictable, seemingly by as much as 12 minutes ahead. This shows that price discovery mainly takes place on the centralized exchanges, whereas price changes on the decentralized exchange tend to trail those on the centralized exchanges. 

\section{How Volatility, Volume, and Liquidity is Distributed over Time}

To codify our notation we denote the one-minute intraday returns and volume by quadruple
subscripts, $y_{w,d,h,m}$, to identify the \emph{week}, \emph{day},
\emph{hour}, and \emph{minute} from which the intraday returns are sampled.
Week numbers are simply $w=1,\ldots,W$ where $W$ is the number of weeks in our sample
period. Day-of-the-week is labelled with $d=1,\ldots,7$ with $d=1$ corresponding to Mondays and $d=2$ to Tuesdays, etc. and hours and minutes are labelled, $h=00,\ldots,23$ and $m=00,\ldots,59$,
respectively. 

To measure patterns in volatility and volume, we introduce
relative measures of volatility and volume across days, hours, and minutes. We measure volatility using absolute returns rather than squared returns. The reason being that absolute returns are less sensitive to "outliers'' and suffice for measuring relative volatility. Absolute returns and the closely related range-based measures are commonly used to measure and model volatility, see e.g. \citet{AlizadehBrandtDiebold02}. Another advantage to using absolute returns is that they, in conjunction with volume, can be used to compute some illiquidity measures we present below. The corresponding results, based on squared returns, are very similar and, as expected, are a bit more noisy. The relative measures gain additional robustness by being averages of variables that are bounded between zero and one. The day-of-the-week measures of volatility and volume measured relative to that over the preceding week is defined by:\footnote{To simplify our notation, we use the following equivalence class notation,
\[
(w,d,h,m)=(w-i,d+7i-j,h+24j-k,m+60k),\qquad\forall i,j,k\in\mathbb{Z}.
\]
This simplifies the expressions for denominators in the relative quantities.} 
\[
\lambda_{\sigma}^{\mathrm{day}}(d)\equiv\frac{1}{N_{d}}\sum_{w}\tfrac{7\sum_{h,m}|y_{\tau(w,d,h,m)}|}{\sum_{i=0}^{6}\sum_{h,m}|y_{\tau(w,d-i,h,m)}|}\quad\text{and}\quad\lambda_{V}^{\mathrm{day}}(d)\equiv\frac{1}{N_{d}}\sum_{w}\tfrac{7\sum_{h,m}V_{\tau(w,d,h,m)}}{\sum_{i=0}^{6}\sum_{h,m}V_{\tau(w,d-i,h,m)}},
\]
respectively, for $d=1,\ldots,7$. Here, $N_{d}\simeq W$ is the number
of observations for day $d$ in our sample (either $N_{d}=W$ or $N_{d}=W-1$). If we take the example with $d=1$ (Mondays), then $\lambda_{\sigma}^{\mathrm{day}}(1)$ and $\lambda_{V}^{\mathrm{day}}(1)$ measures the amount of volatility and volume, respectively, that occur on Mondays relative to other days of the week, where $\lambda=1$ corresponds to the average level of volatility and volume, respectively.

Similarly, the relative measures of volatility and volume for hour-of-the-day are defined by:
\[
\lambda_{\sigma}^{\mathrm{hour}}(h)\equiv\frac{1}{N_{h}}\sum_{w,d}\tfrac{24\sum_{m}|y_{\tau(w,d,h,m)}|}{\sum_{j=0}^{23}\sum_{m}|y_{\tau(w,d,h-j,m)}|}\quad\text{and}\quad\lambda_{V}^{\mathrm{hour}}(h)\equiv\frac{1}{N_{h}}\sum_{w,d}\tfrac{24\sum_{m}V_{\tau(w,d,h,m)}}{\sum_{j=0}^{23}\sum_{m}V_{\tau(w,d-i,h,m)}},
\]
respectively, for $h=0,\ldots,23$, where, $N_{h}\simeq7\times W$
is the number of observations for hour $h$. 

Finally, for minute-of the hour we define volatility and volume relative
to the previous hour by 
\[
\lambda_{\sigma}^{\mathrm{minute}}(m)\equiv\frac{1}{N_{m}}\sum_{w,d,h}\tfrac{60\times |y_{\tau(w,d,h,m)}|}{\sum_{k=0}^{59}|y_{\tau(w,d,h-j,m)}|}\quad\text{and}\quad\lambda_{V}^{\mathrm{minute}}(m)\equiv\frac{1}{N_{m}}\sum_{w,d,h}\tfrac{60\times V_{\tau(w,d,h,m)}}{\sum_{k=0}^{59}V_{\tau(w,d-i,h,m)}},
\]
respectively for $m=0,\ldots,59$, where $N_{m}\simeq7\times24\times W$
is the number of observations for minute $m$.
\begin{figure}[H]
\begin{centering}
\includegraphics[width=0.33\textwidth]{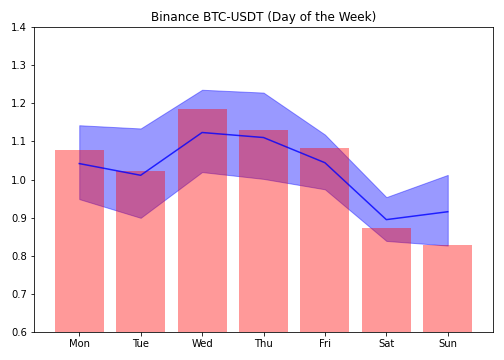}%
\includegraphics[width=0.33\textwidth]{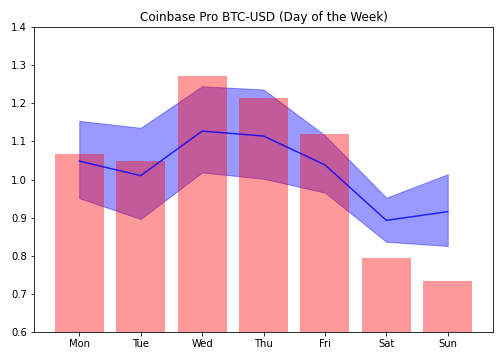}
\includegraphics[width=0.33\textwidth]{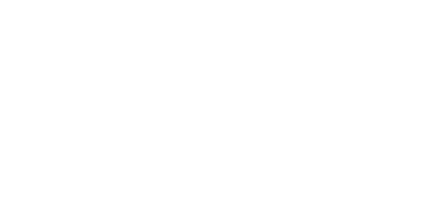}
\par
\includegraphics[width=0.33\textwidth]{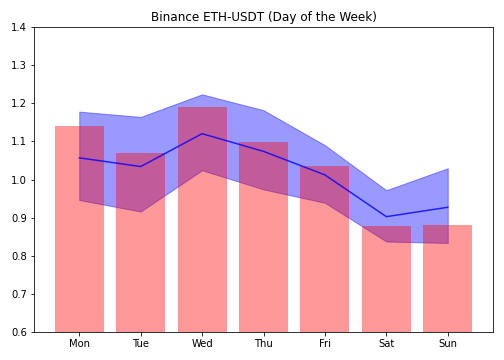}%
\includegraphics[width=0.33\textwidth]{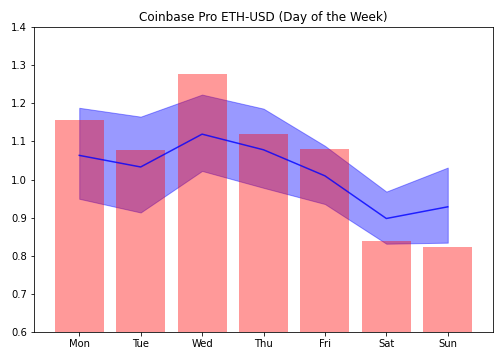}%
\includegraphics[width=0.33\textwidth]{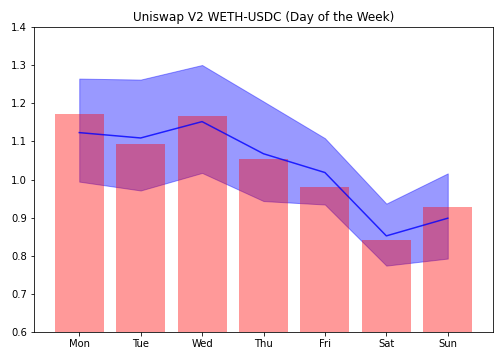}
\par\end{centering}
\begin{small}\caption{Day-of-the-week relative volatility, $\lambda_{\sigma}^{\mathrm{day}}(d)$,
(blue line) and relative volume, $\lambda_{V}^{\mathrm{day}}(d)$,
(red bars). The shaded areas are 95\% confidence bands for the relative
volatility. Results for Bitcoin are shown in upper panels, Ethereum
in lower panels, and results for Binance, Coinbase Pro, and Uniswap V2
are shown in left, middle, and right panels, respectively.\label{fig:Day-of-Week-Relative} }
\end{small}
\end{figure} 

Notice that the ratios in each of the definitions measure the local volatility or volume, relative to a preceding period of time, to account for the persistent variation in volatility over time. 

\subsection{Day-of-Week Patterns}

We present the daily relative measures of volatility (solid blue line) and volume (red bars) in Figure \ref{fig:Day-of-Week-Relative} for each of the five currency pairs. The shaded blue regions display the 95\% confidence intervals for the measure of volatility, and these are relatively large because the number of weeks in the sample period ($N_d\simeq 35$) is relatively small. 
If volatility and/or volume was evenly distributed over the week, we would expect all measures to be about one. However, both volatility and volume tend to be smaller on weekends, which is most pronounced on Saturdays. Note that the relative volume varies more than the relative volatility on the centralized exchanges, which suggests that volume is more concentrated than volatility. 
The uneven distribution of volume and volatility is documented further in the Appendix, see Figure \ref{fig:DayofWeekbyYear}. The same patterns are observed in the extended sample period and are found to be stable over time.
\begin{figure}[H]
\begin{centering}
\includegraphics[width=0.34\textwidth]{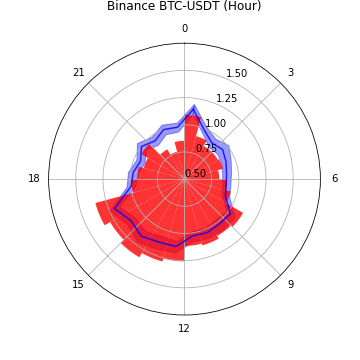}%
\!\!\!%
\includegraphics[width=0.34\textwidth]{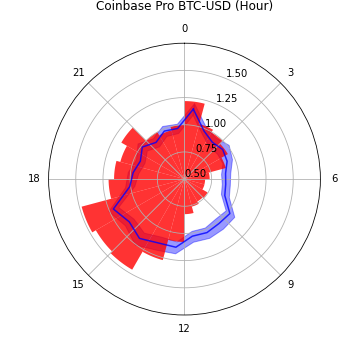}%
\!\!\!%
\includegraphics[width=0.34\textwidth]{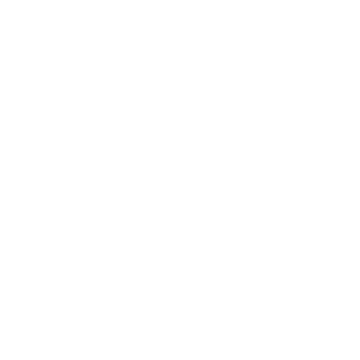}
\par\end{centering}
\begin{centering}
\includegraphics[width=0.34\textwidth]{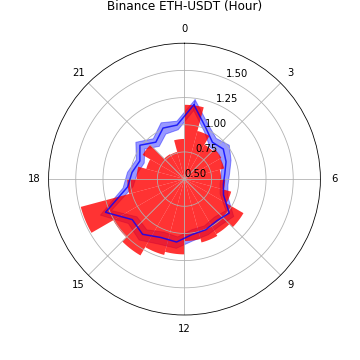}%
\!\!\!%
\includegraphics[width=0.34\textwidth]{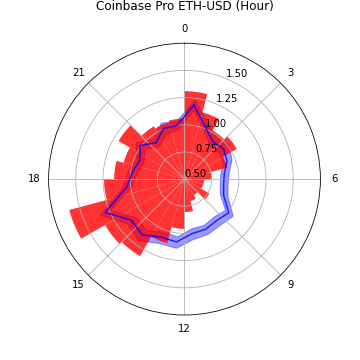}%
\!\!\!%
\includegraphics[width=0.34\textwidth]{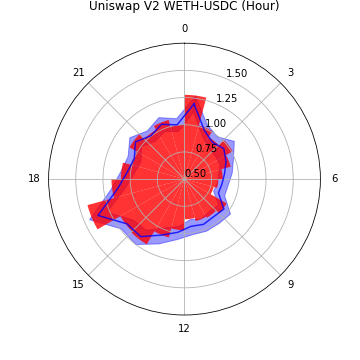}
\par\end{centering}
\begin{small}\caption{Hour-of-the-day Nightingale plots showing hourly volatility (blue line) and volume (red bars) relative to their average levels (normalized to one). The shaded
areas are 95\% confidence bands for the relative hourly volatility.
Results for Bitcoin are shown in the upper panels, Ether in the lower panels,
and results for Binance, Coinbase Pro, and Uniswap V2 are shown in left,
middle, and right panels, respectively.\label{fig:Hour-of-Day-Relative}}
\end{small}
\end{figure}

\subsection{Hours-of-Day Patterns}

For hours-of-the-day we present our results in Figure \ref{fig:Hour-of-Day-Relative}.
We illustrate the patterns in volatility and volume, using radial
line graphs that we refer to as \emph{Nightingale plots}. A Nightingale volatility plot presents the distribution of volatility
over a certain period (such as a day or an hour) using a circular
plot with polar coordinates.\footnote{These polar plots are similar to circular histograms,
and are sometimes called \emph{radar plot}s, \emph{spider plot}s,
\emph{circular histogram}s, and \emph{rose diagram}s. This form of plot was popularized
by Florence Nightingale in 1881.}
Due to arbitrageurs, price and volatility show high levels of agreement between Binance and Coinbase Pro. For hour-of-the-day, WETH/USDC on Uniswap V2 also demonstrates similar intraday volatility patterns as the centralized exchanges, volatility tends to be high around 16 UTC and hit bottom near 5 UTC. \citet{Andersen1998b} also find intraday periodicity in the DM-\$ market, which they attribute to the combined effect of \emph{u}-shape activities in markets in different time zones. 
Unlike volatility, we observe some differences in the patterns of hour-of-the-day volume. These patterns are consistent with the geographical distribution of users. Coinbase Pro has substantially less volume in the hours from 5 UTC to 12 UTC, which coincides with night time in North America. This finding is intuitive because the majority of web-traffic to Coinbase Pro is from the US. The web-traffic to Binance is less concentrated on time-zones, which is consistent with a relatively more even distribution of volume.
Notice that relative volatility and volume have very similar patterns for Uniswap V2. This is a natural consequence of AMM protocols where a price change is directly tied to trading volume (i.e. the amount of a trade determines the price movement along the constant function curve).  This is in contrast to order book based markets where prices can change without trading volume and large price changes can in principle take place with little trading volume. 

Notice the spikes in volume and volatility at 0h and 16h, and less noticeable spikes at 8h. These are observed across all exchanges and for both BTC and ETH. These coincide with funding times in future markets, as we discuss further below.

\subsubsection{Relative Illiquidity Measure for Hour-of-the-Day}
\begin{figure}[H]
\begin{centering}
\includegraphics[width=0.34\textwidth]{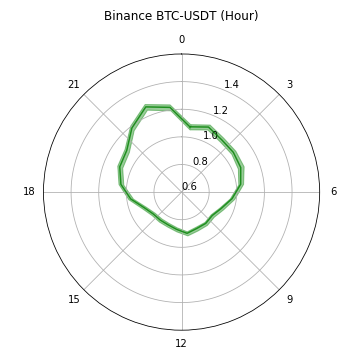}%
\!\!\!%
\includegraphics[width=0.34\textwidth]{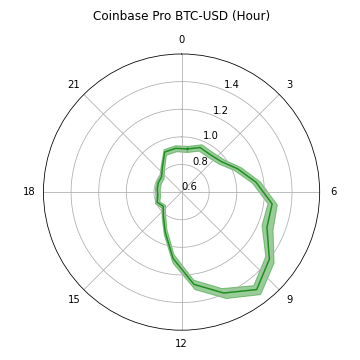}%
\!\!\!%
\includegraphics[width=0.34\textwidth]{figures/Nightingale_Plots/empty}
\par\end{centering}
\begin{centering}
\includegraphics[width=0.34\textwidth]{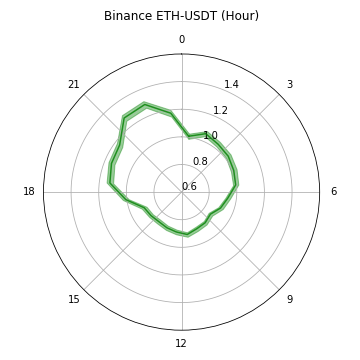}%
\!\!\!%
\includegraphics[width=0.34\textwidth]{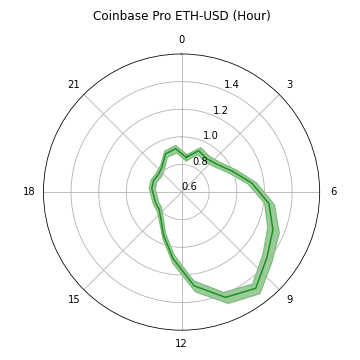}%
\!\!\!%
\includegraphics[width=0.34\textwidth]{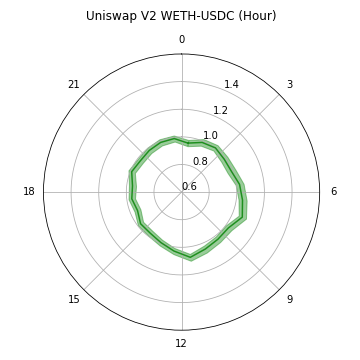}
\par\end{centering}
\begin{small}\caption{Nightingale plots for hourly illiquidity relative to the average level of illiquidity. The shaded
areas are 95\% confidence bands for the relative hourly illiquidity.
Results for Bitcoin are shown in the upper panels, Ether in the lower panels,
and results for Binance, Coinbase Pro, and Uniswap V2 are shown in left,
middle, and right panels, respectively.\label{fig:Hour-of-Day-Illiquidity}}
\end{small}
\end{figure}

We can combine our measurements of volatility and volume into a relative measure of illiquidity based on \cite{Amihud2002}.  First, we compute the illiquidity measure for each hour, and then we compare the hourly measure to the average of the preceding 24 hours. Our relative illiquidity measure is defined by:
\[
\lambda_{\mathrm{Illiquid}}^{\mathrm{hour}}(h)\equiv
\frac{1}{N_{h}}\sum_{w,d}\frac{
    \mathrm{Illiq}(w,d,h)}{
    \tfrac{1}{24}\sum_{j=0}^{23}
    \mathrm{Illiq}(w,d,h-j)}\qquad\text{where}\quad
    \mathrm{Illiq}(w,d,h) \equiv \frac{\sum_{m}|y_{\tau(w,d,h,m)}|}{V_{\tau(w,d,h)}}.
\]
The distribution of illiquidity over the hours of the day is show in Figure \ref{fig:Hour-of-Day-Illiquidity}. 
For Binance, the relative illiquidity is largest during night time in East-Asia where it is about 30\% larger for both Bitcoin and Ether.
For Coinbase Pro, we see the opposite distribution with illiquidity peaking at 10:00 UTC, which is early in the morning in the US. The variation in the liquidity measure is a bit larger for Coinbase, ranging from about 80\% to nearly 150\%. We also observe that the illiquidity patterns for a given exchange, Binance and Coinbase Pro, are nearly identical for the two cryptocurrencies. For Uniswap V2, the liquidity is quite evenly distribution over the 24 hours of the day and never deviates more than 5 percent from the average. This is largely a feature of the design of the decentralized exchanges where changes in prices are directly linked to trading. Aggregating the data for the three exchanges creates a combined measure of illiquidity similar to that of Binance alone because volume on Binance is many times larger, even during the hours where Coinbase Pro has the most trading activity (around 16 UTC). 

The relative illiquidity measures from centralized exchanges are mainly driven by variation in relative volume. The numerators, in their illiquidity measures, are nearly identical because their prices are highly correlated. Therefore, a large value of the relative illiquidity (e.g. the value at 10 UTC for Coinbase Pro) does not imply that trading on Coinbase Pro at this hour is likely to have a larger price impact than other times of the day. 
\begin{figure}[H]
\begin{centering}
\includegraphics[width=1\textwidth]{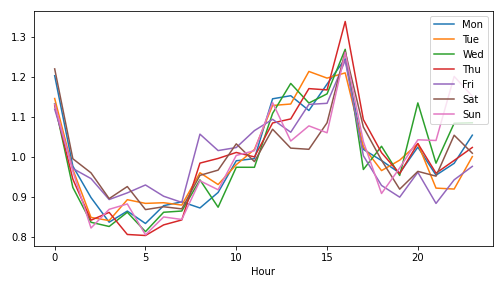}
\par\end{centering}
\begin{small}
\caption{BTC/USDT (Binance) Hour-of-the-Day patterns for each day of the week. Sample period is from Jan 1, 2019 to May 31, 2021. \label{fig:rel-vol-hour-in-each-weekday}}
\end{small}
\end{figure}

\subsubsection{Hour-of-the-Day by Weekday}
We have seen evidence of strong patterns in volatility and volume across weekdays and hours of the day. Some of these patterns could, in principle, be driven by particular weekdays or hours of the day.
For this reason we proceed to investigate if the hour-of-the-day
effect varies by weekday. 
To this end we define the following quantities of relative volatility,
\[
R_{d}(h)=\tfrac{1}{N_h}\sum_{w,d}\frac{\sum_{m=1}^{60} |y_{\tau(w,d,h,m)}|}{\tfrac{1}{24}\sum_{j=0}^{23} \sum_{m=1}^{60}|y_{\tau(w,d,j,m)}^{2}|},\quad\text{for}\quad h=0,\ldots,23,\ d=1,\ldots,7.
\]
This measure is similar to $\lambda_{\sigma}^{\mathrm{hour}}(h)$ but is now computed separately for each of the seven weekdays and the ratio of volatility for the $h$-th hour is now measured relative to the 24 hours within the same day of the week. It is similar to a statistic used in \citet{WangLiuHsu:2020}, albeit they used squared returns and did not separate the hourly share of volatility by day of the week.

Figure \ref{fig:rel-vol-hour-in-each-weekday} shows that the distribution of volatility over the hours of the day are fairly similar across weekdays. The 1st hour and the hour after 16:00 UTC tend to have the highest level of volatility. The ratio of volatility during the first hour of the day is particularly large on Mondays, while the 13th and 15th hours tend to be relatively calmer on Fridays, Saturdays, and Sundays than on other weekdays.
\begin{figure}[H]
\begin{centering}
\includegraphics[width=0.34\textwidth]{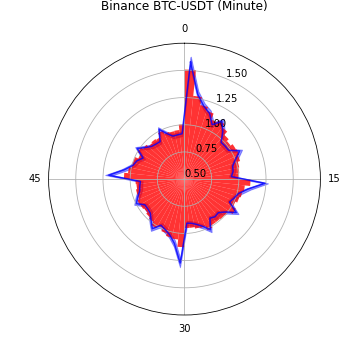}%
\!\!\!%
\includegraphics[width=0.34\textwidth]{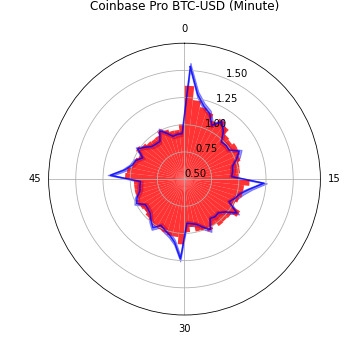}%
\!\!\!%
\includegraphics[width=0.34\textwidth]{figures/Nightingale_Plots/empty}
\par\end{centering}
\begin{centering}
\includegraphics[width=0.34\textwidth]{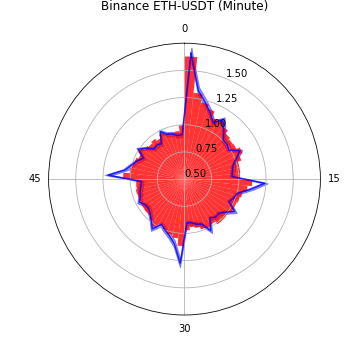}%
\!\!\!%
\includegraphics[width=0.34\textwidth]{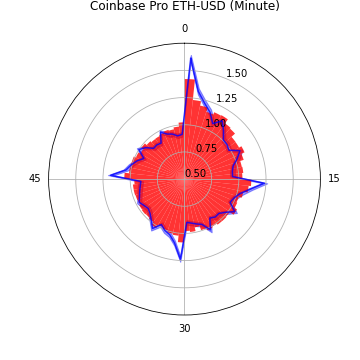}%
\!\!\!%
\includegraphics[width=0.34\textwidth]{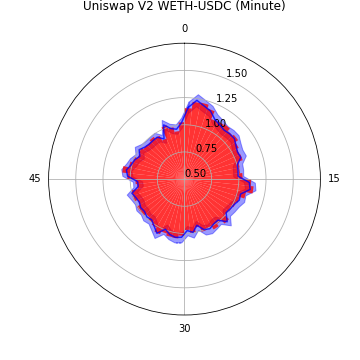}
\par\end{centering}
\begin{small}\caption{Patterns in volatility (blue) and volume (red) by the minute-of-the-hour. Results for Bitcoin are shown in the upper panels, Ether in the lower panels. The data sources are: Binance (left panels), Coinbase Pro (middle panels), and Uniswap V2 (right panels).\label{fig:Minute-of-Hour-Relative}}\end{small}
\end{figure}

\subsection{Periodicity within the Hour}

Next we turn to the minute-of-the-hour analysis. The Nightingale plots are shown in Figure \ref{fig:Minute-of-Hour-Relative}. These plots show very strong minute-of-the-hour patterns on the centralized exchanges. These patterns are most pronounced in volatility. There is a large burst in volatility during the first minute of the hour and medium bursts immediately after minutes 15, 30, and 45, and smaller bursts every five minutes. The most likely explanation for these patterns is algorithmic trading, possibly magnified by individual traders who rely on chart signals.
\begin{figure}[H]
\begin{centering}
\includegraphics[width=0.45\textwidth]{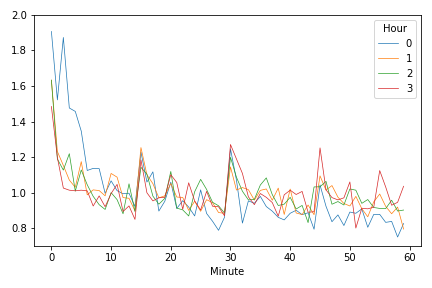}\includegraphics[width=0.45\textwidth]{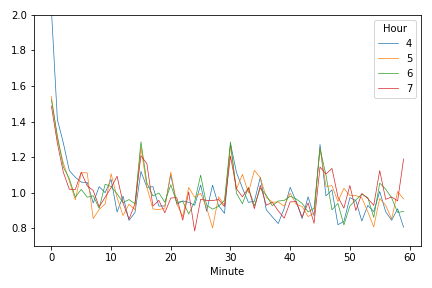}
\par\end{centering}
\begin{centering}
\includegraphics[width=0.45\textwidth]{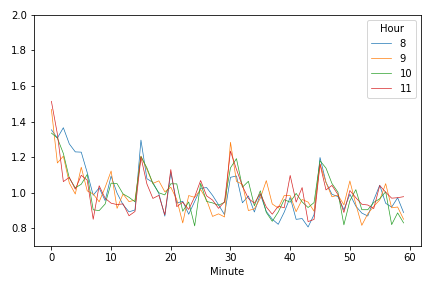}\includegraphics[width=0.45\textwidth]{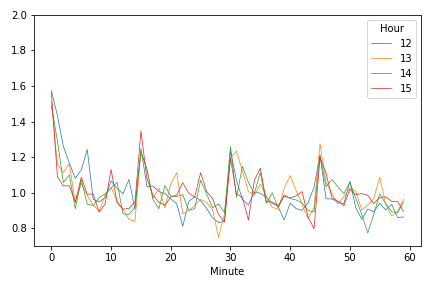}
\par\end{centering}
\begin{centering}
\includegraphics[width=0.45\textwidth]{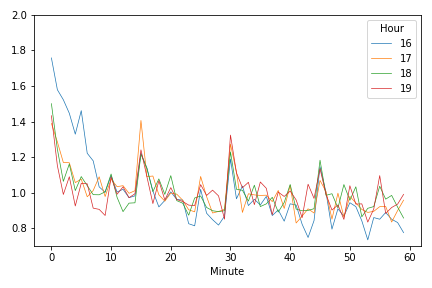}\includegraphics[width=0.45\textwidth]{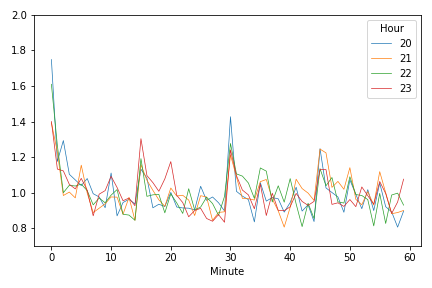}
\par\end{centering}
\begin{small}
\caption{BTC/USDT (Binance) Minute-of-hour patterns in volatility, $R_h(m)$, for each of the 24 hours
of the day. Sample period is from Oct 1, 2020 to May 31, 2021.\label{fig:rel-vol-minute-of-each-hour}}
\end{small}
\end{figure}

We proceed to see if the minute-of-the-hour volatility is similar across hours of the day or more pronounced on certain hours of the day. We define the relative minute-of-the-hour volatility for each of the 24 hours of the day:
\[
R_{h}(m)=\tfrac{1}{N_m}\sum_{w,d,h}\frac{|y_{\tau(w,d,h,m)}|}{\tfrac{1}{60}\sum_{i=0}^{59}|y_{\tau(w,d,h,i)}|}.
\]
This quantity is similar to $\lambda_{\sigma}^{\mathrm{minute}}(m)$. The difference is that  $R_{h}(m)$ measures volatility for the $m$-th minute relative to other minutes within the same hour-of-the-day, whereas $\lambda_{\sigma}^{\mathrm{minute}}(m)$ measures it relatively to the 60 minutes period that just ended. This ensures that the average is one,  $\tfrac{1}{60}\sum_{m=1}^{60} R_{h}(m)=1$, for all $h$. 

Figure \ref{fig:rel-vol-minute-of-each-hour} presents the distribution of volatility over the minutes of the hour, for each of the 24 hours of the day. The patterns we observe in Figure \ref{fig:Minute-of-Hour-Relative} are also observed for each of the 24 hours. The first minute of the hour tends to have the most volatility with peaks every 15 minutes, and smaller peaks every 5 minutes across most hours. 
The minute-of-the-hour effect tends to be most pronounced every fourth hour, 00h,$\ldots$,20h, with the exception of 08h. 
We conjecture that these patterns are related to the funding of perpetual futures. On Binance, the funding takes occur at 00h, 08h, and 16h, on FTX at every hour, and on Bitmex at 04h, 12h, 20h in UTC. 

In Appendix A, we present additional empirical results that show that the patterns are not specific to our sample period. Nearly identical patterns are observed in different calendar years.

Finally, we push the limits of the time-scale by using second-by-second level data for the centralized exchanges. Market microstructure noise will influence the returns at this frequency. However, if this noise is fairly evenly distributed over time, it need not influence our relative measures of volatility and volume. Thus, we compute the distribution of volatility using 
\[
\lambda^{\mathrm{second}}_{\sigma}(s)\equiv
    \tfrac{1}{N_s}\sum_{w,d,h}\frac{|y_{\tau(w,d,h,s)}|}{\tfrac{1}{3600}\sum_{i=0}^{3599}|y_{\tau(w,d,h,s-i)}|}
\] 
where the last argument in $y_{\tau(w,d,h,s)}$ now represents the second within an hour, $s=0,\ldots,3599$ (as opposed to the minute-of-the-hour) and where $N_{s}\simeq7\times24\times W$
is the number of observations for second $s$. The corresponding measure for volume is computed similarly using second-by-second volume data.

We present results for the distribution of volatility and volume over the hour for Bitcoin and Ether using second-by-second data from Binance and Coinbase Pro.
These results are calculated from an extended sample period of Jan 1, 2019 to May 31, 2021 and separate plots are presented for each calendar year. The results are presented in Figure \ref{fig:Second-of-Hour-All}. The results for Binance, BTC/USDT (upper left 6 panels) are also shown in the appendix with a higher resolution, see Figure \ref{fig:Second-of-Hour-Binance}.

The patterns in both volatility and volume are also very clear in the second-by-second data, and some new details emerge at this granularity. For volume, there is a large spike at the beginning of the hour that gradually fades away over several minutes. Only BTC/USD on Coinbase Pro in 2019 does not appear to have a spike in the first second. On Binance, there is another spike of similar magnitude halfway through the hour in 2020 and 2021. On Coinbase Pro, we observe, the puzzling result that the largest spike occurs at the 1200th second, which is 20 minutes into the hour. Albeit, this spike in volume is shorter lived than that at the beginning of the hour. We also notice a smaller burst in volume every minute of the hour. These patterns are consistent with algorithmic trading, which is designed to have intermittent trading activity. Moreover, the patterns in relative volatility and relative volume appear to have grown stronger over time on both exchanges.
\begin{figure}[H]
\begin{centering}
\renewcommand{\tabcolsep}{0pt}
\begin{tabular}{ p{0.02\textwidth}p{0.245\textwidth}p{0.245\textwidth} p{0.245\textwidth} p{0.245\textwidth} }
& \multicolumn{2}{c}{Binance (BTC/USDT)}&\multicolumn{2}{c}{Coinbase Pro (BTC/USD)}\\
\rotatebox{90}{\makebox[0.16\textwidth]{\small 2019}}&
\includegraphics[width=0.245\textwidth]{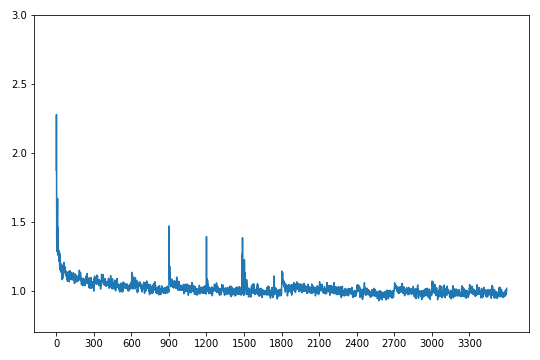}%
&
\includegraphics[width=0.245\textwidth]{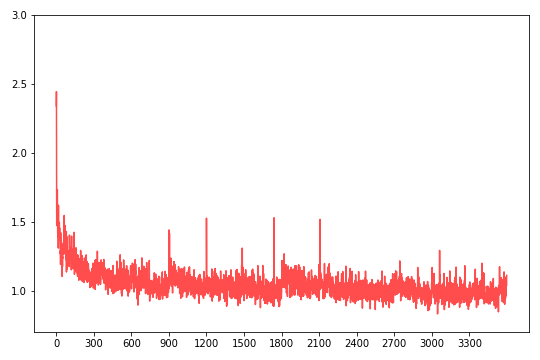}%
&
\includegraphics[width=0.245\textwidth]{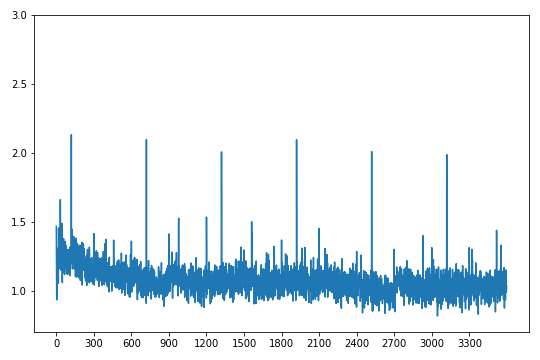}%
&
\includegraphics[width=0.245\textwidth]{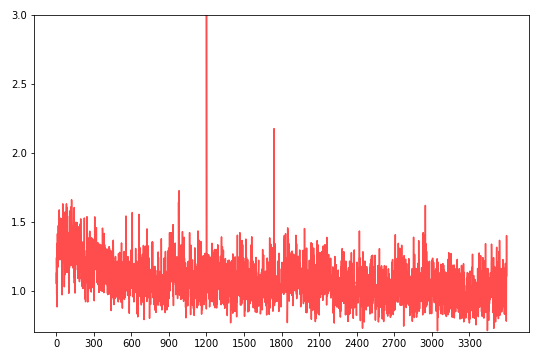}%
\\
\rotatebox{90}{\makebox[0.16\textwidth]{\small 2020}}&
\includegraphics[width=0.245\textwidth]{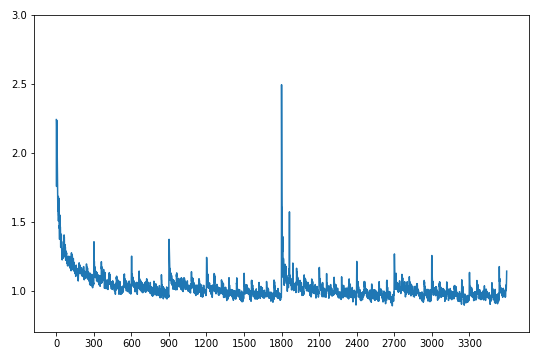}%
&
\includegraphics[width=0.245\textwidth]{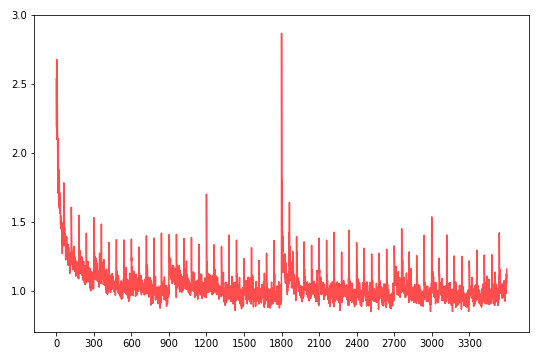}%
&
\includegraphics[width=0.245\textwidth]{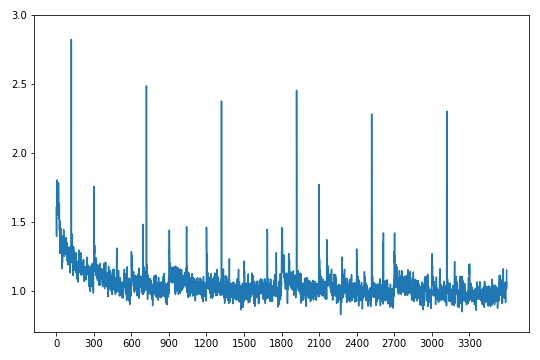}%
&
\includegraphics[width=0.245\textwidth]{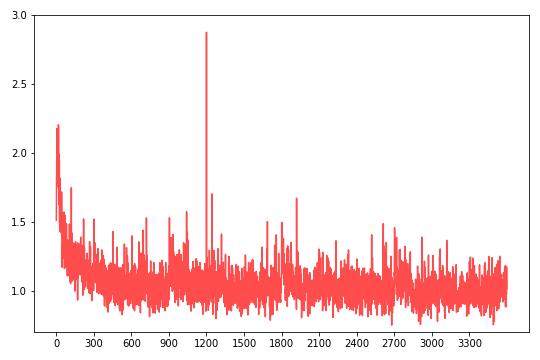}%
\\
\rotatebox{90}{\makebox[0.16\textwidth]{\small 2021}}&
\includegraphics[width=0.245\textwidth]{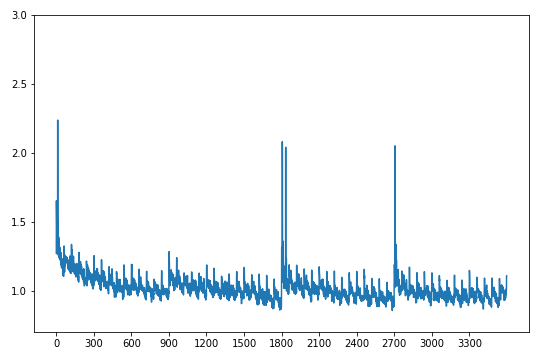}%
&
\includegraphics[width=0.245\textwidth]{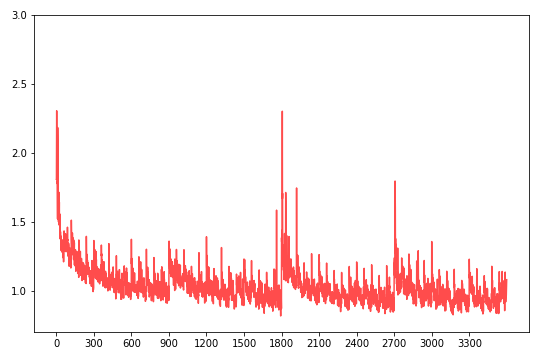}%
&
\includegraphics[width=0.245\textwidth]{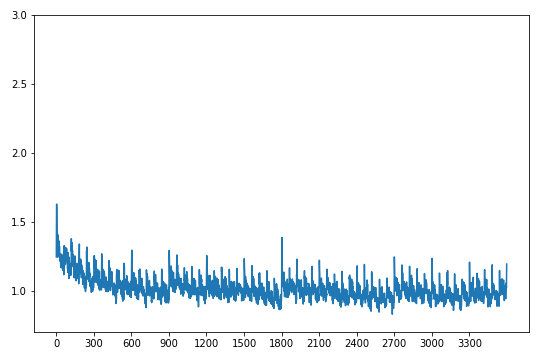}%
&
\includegraphics[width=0.245\textwidth]{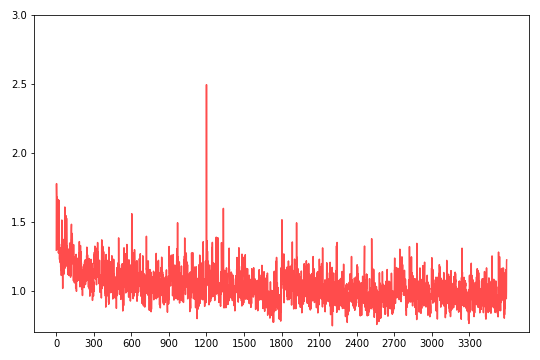}%
\\[3pt]
& \multicolumn{2}{c}{Binance (ETH/USDT)}&\multicolumn{2}{c}{Coinbase Pro (ETH/USD)}\\
\rotatebox{90}{\makebox[0.16\textwidth]{\small 2019}}&
\includegraphics[width=0.245\textwidth]{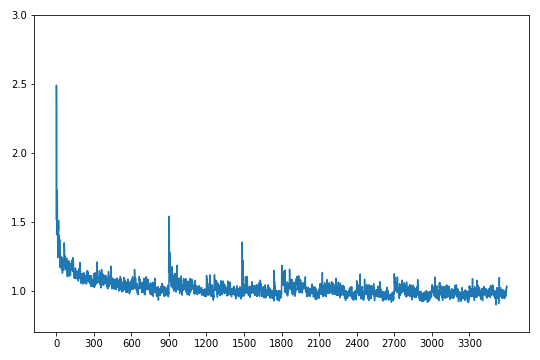}%
&
\includegraphics[width=0.245\textwidth]{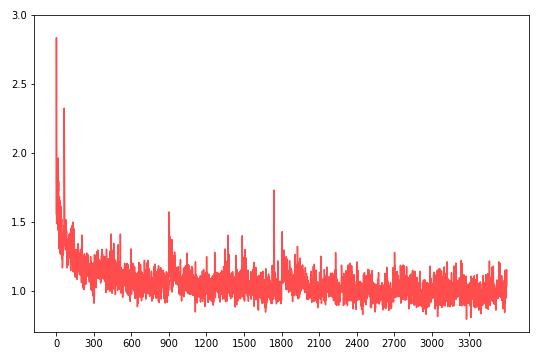}%
&
\includegraphics[width=0.245\textwidth]{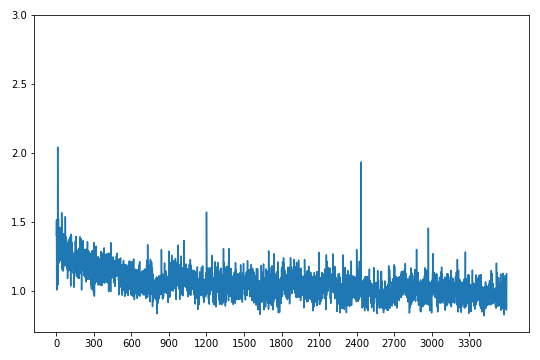}%
&
\includegraphics[width=0.245\textwidth]{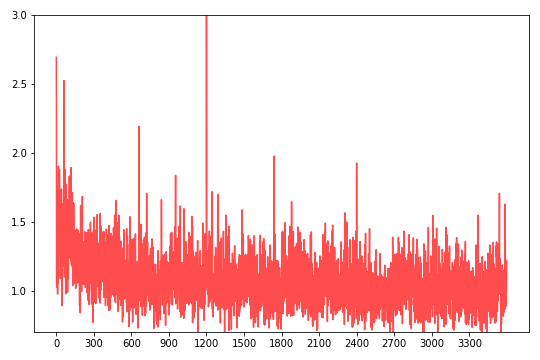}%
\\
\rotatebox{90}{\makebox[0.16\textwidth]{\small 2020}}&
\includegraphics[width=0.245\textwidth]{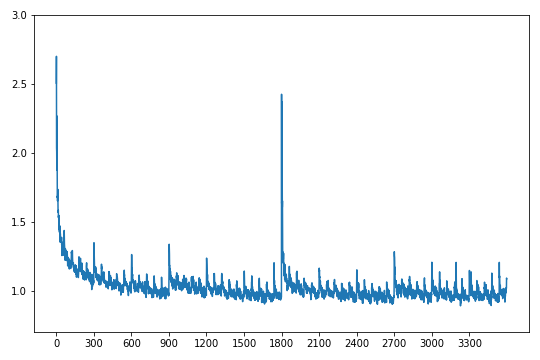}%
&
\includegraphics[width=0.245\textwidth]{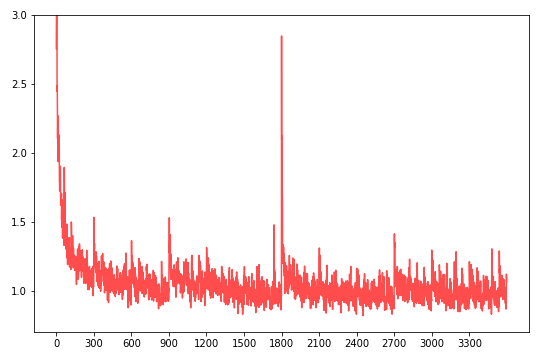}%
&
\includegraphics[width=0.245\textwidth]{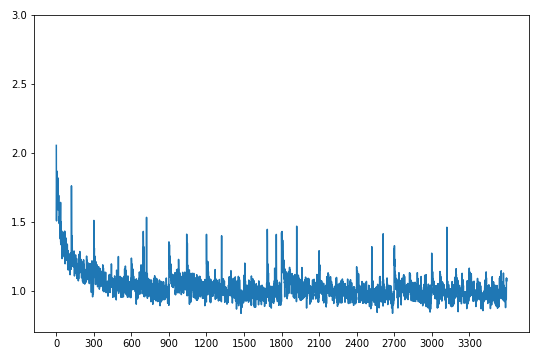}%
&
\includegraphics[width=0.245\textwidth]{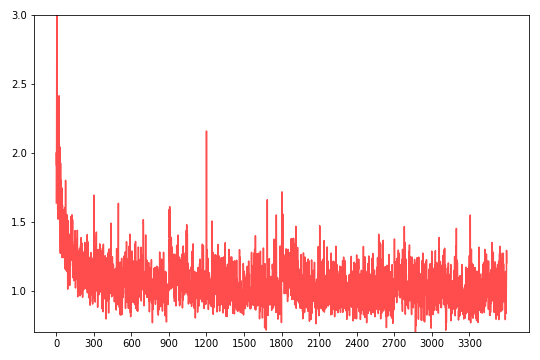}%
\\
\rotatebox{90}{\makebox[0.16\textwidth]{\small 2021}}&
\includegraphics[width=0.245\textwidth]{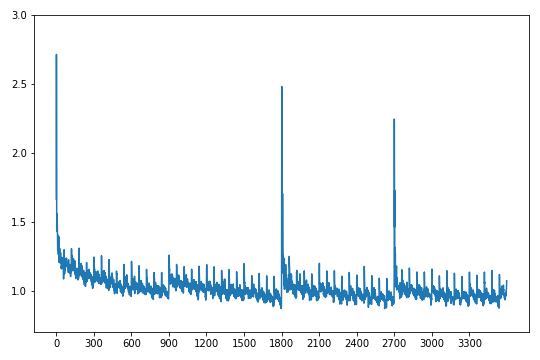}%
&
\includegraphics[width=0.245\textwidth]{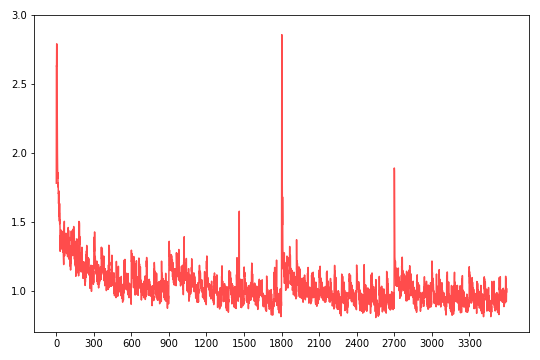}%
&
\includegraphics[width=0.245\textwidth]{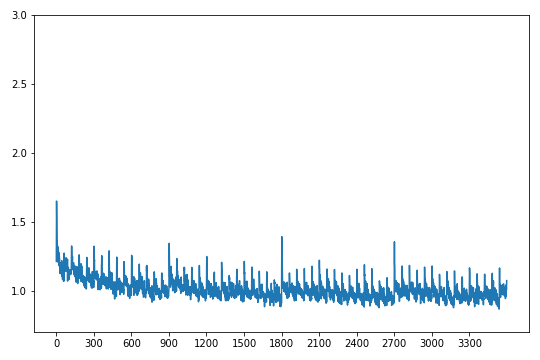}%
&
\includegraphics[width=0.245\textwidth]{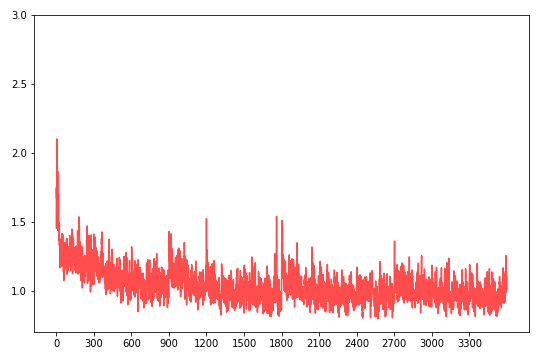}%
\end{tabular}
\renewcommand{\tabcolsep}{6pt}
\par\end{centering}
\begin{small}\caption{Relative volatility and volume by second-of-the-hour for Bitcoin and Ether traded on Binance and Coinbase Pro. The first column of panels present the results for relative volatility on Binance, followed by relative volume on Binance, and the corresponding results for Coinbase Pro. The upper half present the results for Bitcoin and the lower half the results for Ether.\label{fig:Second-of-Hour-All}}\end{small}
\end{figure}

As we discussed in the introduction, periodicity is well documented in exchange rates and stock returns. However this literature has focused on the distribution over the trading day with a continuous distribution (typically U-shaped), as opposed to the within-the-hour periodicity we have documented for cryptocurrencies in this paper. In unreported results, we found very similar and significant within-hour patterns in volatility for the S\&P500 index, so the within-hour patterns do not appear to be specific to cryptocurrencies. In terms of US stock markets, \citet{BroussardNikiforov:2014} documented similar patterns within the hour for traded volume and the number of transactions. The authors did not comment on patterns in volatility, but such can be inferred indirectly from their reported absence of patterns in illiquidity measures. 

\section{Periodicity in Volatility Modeling and Forecasting}
In this section, we show that it is beneficial to account for periodicity in volatility modeling and volatility forecasting. We estimate five GARCH models and compare the empirical fit across models with and without day-of-the-week periodicity. Conventional GARCH models typically need several years of daily returns to produce reliable estimates and, because we divide our sample into an in-sample and an out-of-sample period, we use an extended sample period from January 1, 2017 to July 31, 2021. This data is solely from Coinbase Pro because data from Binance and Uniswap V2 are not available for the entirety of this sample period.

Let ${r_t}=100(\log P_t -\log P_{t-1})$ denote a time series of daily returns in percent (BTC/USD or ETH/USD). A day spans the 24 hour period starting at 0:00h UTC, and we denote the conditional variance by, $h_t=\mathrm{var}(r_t|\mathcal{F}_{t-1})$ where $r_t$ is adapted to the filtration, $\mathcal{F}_{t}$.

We estimate the GARCH model by \cite{bollerslev:86} and the EGARCH model by \cite{Nelson91}. These are two of many GARCH models, see \cite{HansenLundeBeatGarch} and \cite{Bollerslev:10}, and the variation across models is primarily in their specification for the \emph{GARCH equation} that prescribes how $h_t$ evolves over time. The GARCH model has a linear expression for $h_t$, whereas the EGARCH has a linear expression for $\log h_t$. The EGARCH can also accommodate an asymmetric relationship between return shocks and volatility shocks. We also estimate an EGARCH-X model that includes the realized variance as a predictor of future conditional variance, see \cite{Engle2002b}.

A GARCH(1,1) model has 
\[
h_{t+1} = \omega +\beta h_t + \alpha (r_t-\mu)^2,
\]
where $\mu=\mathbb{E}(r_t|\mathcal{F}_{t-1})$. The EGARCH model has the GARCH equation:\footnote{We have simplified the original expression, which included $\alpha(|z_t|-\mathbb{E}|z_t|)$, by letting $\omega$ absorb the constant term $-\alpha\mathbb{E}|z_t|$.}
\[
\log h_{t+1} = \omega +\beta \log h_t +\alpha |z_t| + \tau z_t,
\]
where $z_t=(r_t-\mu)/\sqrt{h_t}$ is the standardized return that has zero conditional mean and conditional variance $\mathrm{var}(z_t|\mathcal{F}_{t-1})=1$.
The EGARCH-X includes the realized variance in the GARCH equation,
\[
\log h_{t+1} = \omega +\beta \log h_t + \gamma\log\tfrac{\mathrm{RV}_t}{h_t} + \alpha|z_t| + \tau z_t,
\]
where the realized variance, $\mathrm{RV}_t$, is computed with five-minute returns we defined in (\ref{eq:RV}). 

We generalize the two EGARCH models to incorporate day-of-the-week periodicity.\footnote{Our structure is related to the Component GARCH by \citet{EngleLee:1999} and the periodic GARCH model by \citet{BollerslevGhysels1996}} The periodic models have a single latent volatility variable, $h^\ast_t$, and the conditional variance for each of the weekdays are simply scaled versions of $h^\ast_t$ to adjust for the variation of relative volatility over day-of-the-week. Specifically,
\[\log h_t=\lambda_{d(t)}+ \log h_t^\ast,\qquad \text{with}\qquad \prod_{d=1}^7 \exp(\lambda_d) = 1,  \]
where $d(t)$ is the mapping from $t$ to the corresponding weekday, $d\in\{1,\ldots,7\}$. Hence, if $t=1$ is a Sunday, then $d(1)=7$  and $d(2)=1$ etc.  
If there is no periodicity, then we have $\lambda_d=0$ for all $d$, whereas $\lambda_d<0$ if weekday $d$ tends to have below average level volatility (as one might expect for Saturdays and Sundays) and $\lambda_d>0$ if weekday $d$ has above average volatility.

When estimating the periodic EGARCH and EGARCH-X models, we reparametrize the model to simplify the estimation problem. Specifically, we estimate a model with, $\log h^\star _{t+1} = \beta \log h^\star _t + \log\tfrac{\mathrm{RV}_t}{h_t} + \alpha|z_t| + \tau z_t $ and $\log h_t=\log h^\star_t + \kappa_{d(t)}$ where $\kappa_d$, $d=1,\ldots,7$, are unrestricted parameters to be estimated. Since $\log h^\ast_t = \log h^\star_t + \omega/(1-\beta)$, we can recover the original parameters using $\omega=\tfrac{1-\beta}{7}\sum_{d=1}^7\kappa_d$ and
$\lambda_d = \kappa_d - \frac{\omega}{1-\beta}$, for $d=1,\ldots,7.$

\begin{table}[ht]
\begin{small}
\begin{tabularx}{\textwidth}{lrrrrrcrrrrr}
\hline
\\[-5pt]
    & \multicolumn{5}{c}{BTC/USD} &  & \multicolumn{5}{c}{ETH/USD}  \\[5pt] 
        \cline{2-6} \cline{8-12}  \\[-5pt]
    &  GARCH  & \multicolumn{2}{c}{EGARCH} & \multicolumn{2}{c}{EGARCH-X}
    && GARCH  & \multicolumn{2}{c}{EGARCH} & \multicolumn{2}{c}{EGARCH-X}    \\[5pt]
$\mu$    & $\underset{(0.112)}{0.181}$ & $\underset{(0.161)}{0.187} $ & $\underset{(0.005)}{0.170}$  
    &  $\underset{(0.113)}{0.174}$ & $\underset{(0.006)}{0.190} $ 
    && $\underset{(0.082)}{0.144}$ & $\underset{(0.161)}{0.187}$ & $\underset{(0.149)}{0.236}$  
    &  $\underset{(0.016)}{0.164}$  & $\underset{(0.150)}{0.223}$  \\
$\beta$  & $\underset{(0.050)}{0.795}$  & $\underset{(0.050)}{0.871}$   & $\underset{(0.031)}{0.896}$  
    &  $\underset{(0.041)}{0.864}$   & $\underset{(0.038)}{0.869}$   
    && $\underset{(0.100)}{0.674}$   & $\underset{(0.050)}{0.872}$    & $\underset{(0.038)}{0.894}$   
    & $\underset{(0.060)}{0.849}$     & $\underset{(0.048)}{0.878}$   \\
$\omega$ & $\underset{(0.618)}{1.448}$    & $\underset{(0.107)}{0.124}$   & $\underset{(0.092)}{0.092}$    
    &  $\underset{(0.143)}{0.334}$   & $\underset{(0.133)}{0.306}$     
    &  & $\underset{(2.212)}{5.198}$   & $\underset{(0.145)}{0.243}$    & $\underset{(0.107)}{0.167}$    
    &  $\underset{(0.206)}{0.396}$   & $\underset{(0.173)}{0.285}$     \\
$\alpha$ & $\underset{(0.037)}{0.131}$    & $\underset{(0.059)}{0.263}$   & $\underset{(0.053)}{0.279}$    
    &  $\underset{(0.110)}{0.120}$     & $\underset{(0.098)}{0.129}$     
    && $\underset{(0.054)}{0.171}$    & $\underset{(0.069)}{0.272}$    & $\underset{(0.058)}{0.265}$     
    &  $\underset{(0.087)}{0.199}$    & $\underset{(0.078)}{0.202}$     \\
$\tau$   &     & $\underset{\phantom{-}(0.033)}{-0.025}$    & $\underset{\phantom{-}(0.031)}{-0.025}$    
    &  $\underset{\phantom{-}(0.036)}{-0.033}$ & $\underset{\phantom{-}(0.034)}{-0.034}$ 
    && & & $\underset{(0.027)}{0.022}$  & $\underset{(0.03)}{0.019}$   & $\underset{(0.028)}{0.026}$ \\
$\gamma$ & & & & $\underset{(0.047)}{0.080}$  &  $\underset{(0.038)}{0.085}$
    && & &  & $\underset{(0.049)}{0.056}$ & $\underset{(0.048)}{0.048}$  \\
$\lambda_{\mathrm{Mo}}$ &&& $\underset{(0.145)}{0.077}$ && $\underset{(0.158)}{0.059}$  
    &&&& $\underset{(0.135)}{0.024}$  && $\underset{(0.138)}{0.016}$  \\
$\lambda_{\mathrm{Tu}}$  &&& $\underset{(0.205)}{0.263}$  && $\underset{(0.233)}{0.220}$  
    &&&& $\underset{(0.147)}{0.240}$  && $\underset{(0.151)}{0.224}$   \\
$\lambda_{\mathrm{We}}$  &&& $\underset{(0.176)}{0.130}$  && $\underset{(0.177)}{0.160}$   
    &&&&$\underset{(0.171)}{0.080}$    && $\underset{(0.172)}{0.098}$   \\
$\lambda_{\mathrm{Th}}$  &&& $\underset{(0.141)}{0.317}$  && $\underset{(0.144)}{0.313}$   
    &&&& $\underset{(0.160)}{0.243}$   && $\underset{(0.162)}{0.241}$  \\
$\lambda_{\mathrm{Fr}}$  &&& $\underset{\phantom{-}(0.144)}{-0.041}$  && $\underset{(0.150)}{0.008}$   
    &&&& $\underset{(0.157)}{0.101}$   && $\underset{(0.158)}{0.110}$  \\
$\lambda_{\mathrm{Sa}}$    &&& $\underset{\phantom{-}(0.160)}{-0.319}$  && $\underset{\phantom{-}(0.161)}{-0.311}$  
    &&&& $\underset{\phantom{-}(0.153)}{-0.471}$  && $\underset{\phantom{-}(0.153)}{-0.461}$  \\
$\lambda_{\mathrm{Su}}$    &&& $\underset{\phantom{-}(0.160)}{-0.426}$  && $\underset{\phantom{-}(0.167)}{-0.461}$  
    &&&& $\underset{\phantom{-}(0.140)}{-0.218}$  && $\underset{\phantom{-}(0.146)}{-0.227}$ \\[15pt] 
\cline{2-6} \cline{8-12} \\
$\ell_\mathrm{is}$  & -3090.6 & -3090.4 & -3071.6 & -3085.8 & -3066.6 
    &&                -3407.3 & -3407.9 & -3392.5 & -3405.9 & -3390.9 \\[5pt]
$\ell_\mathrm{os}$  & -1666.1 & -1668.1 & -1625.7 & -1649.5 & -1615.8 
    &&                -1817.2 & -1817.9 & -1805.7 & -1810.6 & -1800.9  \\[5pt] 
\hline
\end{tabularx}
\end{small}
\begin{small}\caption{Estimated parameters and the corresponding robust standard errors in parentheses for each of the GARCH models. Parameters are estimated with the in-sample period, January 1, 2017 to December 31, 2019, where $\ell_{\mathrm{is}}$ is the maximized log-likelihood. The log-likelihoods are also evaluated out-of-sample (January 1, 2020 to July 31, 2021) using the in-sample estimates resulting in the out-of-sample log-likelihood denoted $\ell_{\mathrm{os}}$.\label{tab:GARCHmodels}}
\end{small}
\end{table}

The estimated models are presented in Table \ref{tab:GARCHmodels}. The models are estimated by maximizing the in-sample log-likelihood that spans the first three years of data, January 1, 2017 to December 31, 2019. The remaining period, January 1, 2020 to July 31, 2021, is used as an out-of-sample evaluation period. We use a Gaussian specification, $z_t\sim iid N(0,1)$ for which the log-likelihood for day $t$ is $\ell_t=-\tfrac{1}{2}(\log h_t +z_t^2+\log(2\pi))$. The maximized log-likelihood using the specification is denoted $\ell_\mathrm{is}=\sum \ell_t$, where the sum is taken over days in the in-sample period and the corresponding out-of-sample log-likelihood is denoted  $\ell_\mathrm{os}=\sum \ell_t$, where the sum is taken over the out-of-sample period. The standard GARCH and EGARCH model have very similar log-likelihoods for both cryptocurrencies both in-sample and out-of-sample. Adding the realized variance to the GARCH equation improves the empirical fit both in-sample and out-of-sample. This is particularly the case for BTC/USD returns where the in-sample log-likelihood improves by about 5 units and the out-of-sample log-likelihood by as much as 18 units. However, even larger gains in the empirical fit are achieved by accounting for periodicity in the model. For BTC/USD, the periodic model improves the log-likelihood by about 19 units in-sample and about 34 units out-of-sample. For ETH/USD, the improvements are roughly 15 units and 11 units, respectively.
\begin{figure}[H]
\begin{centering}
\includegraphics[width=0.8\textwidth]{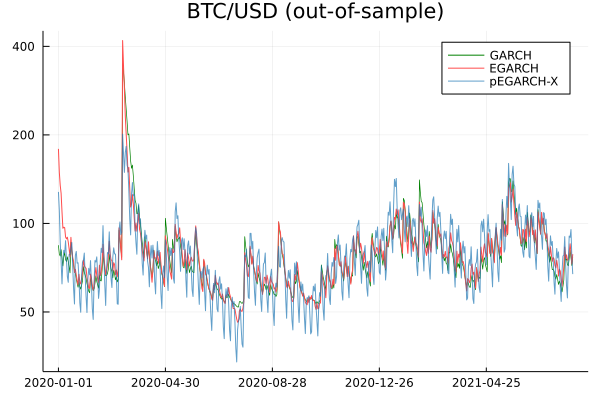}%
\par
\includegraphics[width=0.8\textwidth]{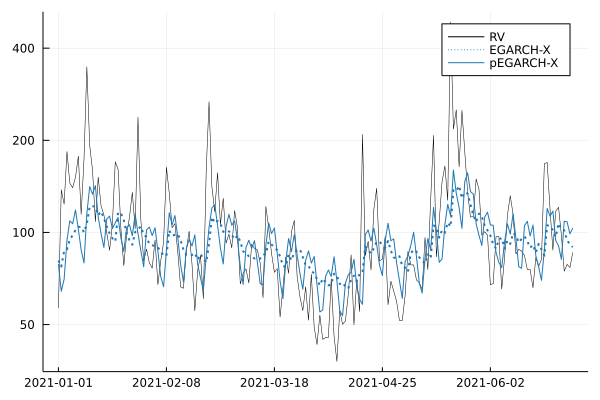}%
\par\end{centering}
\begin{small}\caption{The upper panel displays the conditional variances, $\hat{h}_t$, over the entire out-of-sample period, January 1, 2020 to July 31, 2021 for three models, GARCH, EGARCH, and pEGARCH-X, estimated with daily BTC/USD returns.
The lower panel shows $\hat{h}_t$ for the EGARCH-X model with and without periodicity over the first six months of 2021 (a subset of the out-of-sample period) and the variation in $\mathrm{RV}_t$ over the same period. All measured in annualized volatility.
\label{fig:GARCH_oos}}
\end{small}
\end{figure}

There are several interesting observations to be made from Table \ref{tab:GARCHmodels}. First, the asymmetry parameter, $\tau$, is insignificant in all models. This parameter is typically found to be negative and significant when EGARCH models are estimated with stock returns, see \cite{HansenLundeBeatGarch} and \cite{HansenHuangShek:2012}. For exchange rate data, the GARCH performs on par with the EGARCH, see \cite{HansenLundeBeatGarch}, and a similar situation emerges here with cryptocurrencies because the asymmetry parameter, $\tau$, is insignificant. Second, the models with periodicity show that volatility is below average during the weekend and above average during most weekdays. This variation over the weekdays is similar to that reported in Figure \ref{fig:Day-of-Week-Relative}, which is estimated for a different sample period. There are some noticeable differences such as Wednesday and Fridays, which have a larger share of volatility in Figure \ref{fig:Day-of-Week-Relative} than suggested by the estimates in Table \ref{tab:GARCHmodels}. This could suggest that the distribution of volatility across weekdays has shifted over time. For this reason, it might be possible to develop even better volatility models for cryptocurrencies in which the share of volatility for different weekdays is time varying similar to the time variation in the level of volatility. We shall not pursue this task here because our objective is merely to demonstrate that volatility modeling and forecasting can be improved by taking periodicity into account. 

The out-of-sample volatility paths produced by three estimated GARCH models are shown in the upper panel of Figure \ref{fig:GARCH_oos}. The volatility displayed is in units of annualized volatility, defined by $\sqrt{365\hat{h}_t}$, for each of the three models. The two models without periodicity, GARCH (green line) and EGARCH (red line), produce very similar volatility paths, whereas the EGARCH-X with periodicity has a clear weekly pattern in the volatility series. The lower panel has the annualized volatility for the two variants of the EGARCH-X model for the first six months of 2021 as well as the daily realized variance computed with 5-minute returns. The panel shows that the model with periodicity is in better agreement with the realized variance than the EGARCH-X model without periodicity, which is unable to generate the sort of weekly variation seen in the realized variance.  
The estimated series for  $\hat{h}_t$ for additional models and for both cryptocurrencies, BTC/USD and ETH/USD, are presented in Figure \ref{fig:GARCHmodels}  in the appendix.

\section{Concluding Remarks}
We have documented intra-week, intra-day, and intra-hour patterns in volatility and volume for two leading cryptocurrencies and for three exchanges. We believe we are the first to document the patterns within the hour, which appear to have grown stronger in recent years. Interestingly, the within the hour periodicity does not appear to be specific to cryptocurrencies because we have found the same patterns in volatility for the S\&P 500 index in unreported results. In future research, it will be interesting to determine if these patterns are primarily driven by algorithmic trading or have other explanations.

We have also noted some important differences between centralized and decentralized exchanges. The relatively high cost of transacting on Uniswap V2 causes prices to be sluggish. This cost entails the protocol fee and the network's cost of gas, which, when paired with slippage, often negates or greatly reduces arbitrage opportunities. This is evident from the cross-correlations in returns from different exchanges and explains that the within the hour patterns are much less pronounced on Uniswap V2. An implication of these empirical findings is that price discovery takes place on the centralized exchanges, and the higher levels of autocorrelation in returns from the decentralized exchange makes them less suitable for measuring volatility. 

Our empirical findings are important for volatility modeling, volatility measurement, volatility forecasting, and the interpretations of realized measures of volatility. For instance, the equation that specifies the dynamic properties of volatility in GARCH models and stochastic volatility models should reflect the fully anticipated variation in relative volatility over the week. We have shown that GARCH models with day of the week periodicity can lead to better forecasting performance. Since we found the periodicity within the day and within the hour to be even stronger than the variation across weekdays, it is reasonable to expect that models of volatility at higher frequencies than daily returns would benefit from incorporating periodicity. 

The intra-hour periodicity could also have important implications for some realized measures of volatility, such as the bipower variation estimator by \citet{barndorff-shephard:2004BiPower} and the nearest neighbor truncation estimator by \citet{AndersenDobrevSchaumburg2012}. These estimators rely on adjacent intraday returns being approximately homogeneous. However, the first 1-second intraday return of each hour and the preceding 1-second return have very different properties. Similar discontinuities are observed every 15th minute, every 5th minute,  and (to a lesser extend) every one minute. This inhomogeneity carries over to intraday returns with lower sampling frequencies in calendar time, such as 1-minute returns. It is likely possible to lessen the inhomogeneity by sampling more frequently when volatility is expected to be high, and less frequently when it is expected to be low. The type of sampling was coined \emph{business time sampling} by \citet{Oomen2006}.

Periodicity is important for the interpretation of changes in real-time measures of volatility and volume. For instance, an observed change in volatility may signal an unanticipated change in the level of volatility, and this may be important new information. However, if the change is fully anticipated because it follows the pattern of volatility that is typically seen at this time, then the change in volatility does not reflect new information. Therefore, the usefulness of information contained in a measurement is not the measurement itself, but is the difference between the actual measurement and the expected measurement, and periodicity is an important component of the latter.

{\footnotesize{}\bibliographystyle{agsm}
\bibliography{prh}
}{\footnotesize\par}

\clearpage
\appendix
\setcounter{equation}{0}\renewcommand{\theequation}{A.\arabic{equation}}
\setcounter{figure}{0}\renewcommand{\thefigure}{A.\arabic{figure}}
\setcounter{table}{0}\renewcommand{\thetable}{A.\arabic{table}}

\section{Supplementary Empirical Results}

We present summary statistics for the sample period, October 1, 2020 to May 31,
2021, in Table \ref{tab:Summary}. The corresponding summary statistics for the extended sample period, January 1, 2019 to May 31, 2021 are presented in Table \ref{tab:SummaryLongSample}. The Table does not include statistics for Uniswap V2, because  it only launched in May 2020.

\begin{table}[H]
\centering{}\begin{small} 
\begin{tabular*}{1\textwidth}{@{\extracolsep{\fill}}>{\raggedright}p{0.26\textwidth}rrrrrr} 
\hline \\[-1pt]
\multicolumn{7}{c}{Volatility RV (5-minute returns)} \\[4pt] 
&\multicolumn{1}{c}{Average}&\multicolumn{1}{c}{min}&\multicolumn{1}{c}{q25}&\multicolumn{1}{c}{q50}&\multicolumn{1}{c}{q75}& \multicolumn{1}{c}{max}\\[3pt] 
BTC/USDT (Binance)&\multicolumn{1}{c}{ 67.72} & \multicolumn{1}{c}{14.96} & \multicolumn{1}{c}{38.83} & \multicolumn{1}{c}{54.57} & \multicolumn{1}{c}{80.01    } & \multicolumn{1}{c}{635.23}\\[2pt] 
BTC/USD (Coinbase Pro)&\multicolumn{1}{c}{ 69.05} & \multicolumn{1}{c}{11.25} & \multicolumn{1}{c}{38.43} & \multicolumn{1}{c}{55.48} & \multicolumn{1}{c}{82.45    } & \multicolumn{1}{c}{628.57}\\[2pt] 
ETH/USDT (Binance)&\multicolumn{1}{c}{86.29} & \multicolumn{1}{c}{22.52} & \multicolumn{1}{c}{53.10} & \multicolumn{1}{c}{71.59} & \multicolumn{1}{c}{100.78}  & \multicolumn{1}{c}{773.62}\\[2pt]  
ETH/USD  (Coinbase Pro)&\multicolumn{1}{c}{87.81} & \multicolumn{1}{c}{22.64} & \multicolumn{1}{c}{54.04} & \multicolumn{1}{c}{72.77} & \multicolumn{1}{c}{101.63}  & \multicolumn{1}{c}{753.04}\\[2pt] 
\tabularnewline   &  &  &  &  &  & \tabularnewline  \multicolumn{7}{c}{Volume} \\[4pt]   
& \multicolumn{1}{c}{Average} & \multicolumn{1}{c}{min} & \multicolumn{1}{c}{q25} & \multicolumn{1}{c}{q50} & \multicolumn{1}{c}{q75} & \multicolumn{1}{c}{max}\\[3pt] 
BTC/USDT (Binance)& 1,162M & 55M & 282M  & 512M  & 1,099M & 13,481M \\[2pt] 
BTC/USD  (Coinbase Pro)&   312M & 10M & 63M  & 118M  & 307M   &  4,177M \\[2pt] 
ETH/USDT (Binance)&   573M & 16M & 60M   & 138M  & 408M   & 11,645M \\[2pt] 
ETH/USD  (Coinbase Pro)&   172M &  3M & 15M  & 32M   & 88M    &  4,510M \\
&  &  &  &  &  &  \\  
\hline 
\end{tabular*}
\end{small}
\begin{small}\caption{Summary Statistics for the sample period January 01, 2019 to May 31,
2021.\label{tab:SummaryLongSample}}
\end{small}
\end{table}

\begin{figure}[H]
\begin{centering}
\includegraphics[width=0.5\textwidth]{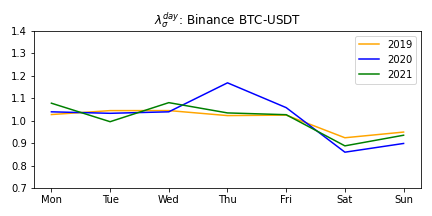}\includegraphics[width=0.5\textwidth]{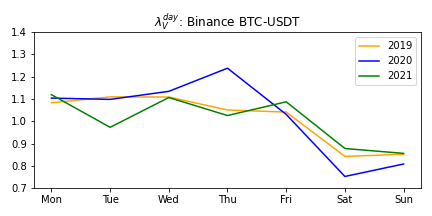}
\par\end{centering}
\begin{centering}
\includegraphics[width=0.5\textwidth]{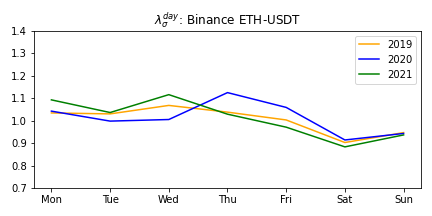}\includegraphics[width=0.5\textwidth]{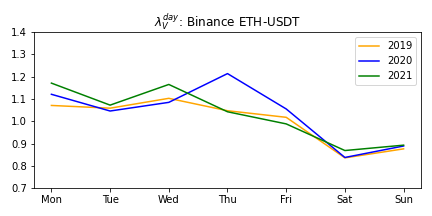}
\par\end{centering}
\begin{centering}
\includegraphics[width=0.5\textwidth]{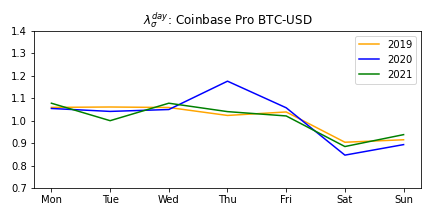}\includegraphics[width=0.5\textwidth]{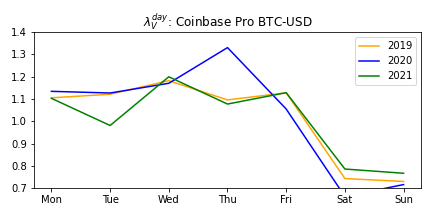}
\par\end{centering}
\centering{}\includegraphics[width=0.5\textwidth]{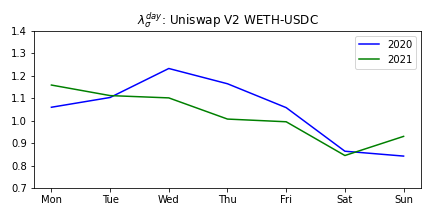}\includegraphics[width=0.5\textwidth]{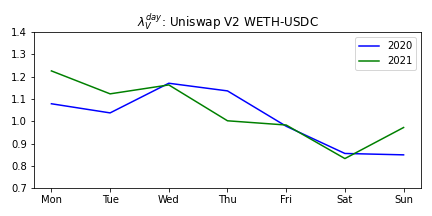}
\begin{small}
\caption{Day-of-the-Week relative volatility (left panels) and relative volume (right panels) by calendar year. The sample period for Binance and Coinbase Pro is  January 1, 2019 to May 31, 2021 
and the sample period for Uniswap V2 is October 1, 2020 to May 31,
2021.\label{fig:DayofWeekbyYear}}
\end{small}
\end{figure}

\begin{figure}[H]
\begin{centering}
\includegraphics[width=0.5\textwidth]{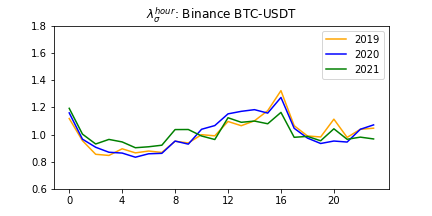}\includegraphics[width=0.5\textwidth]{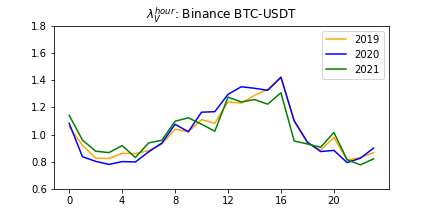}
\par\end{centering}
\begin{centering}
\includegraphics[width=0.5\textwidth]{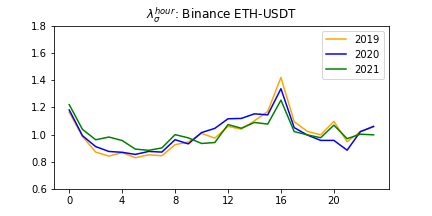}\includegraphics[width=0.5\textwidth]{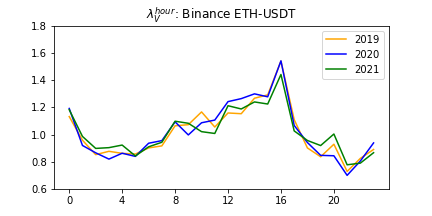}
\par\end{centering}
\begin{centering}
\includegraphics[width=0.5\textwidth]{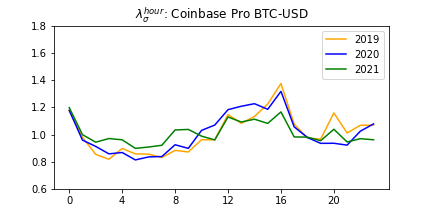}\includegraphics[width=0.5\textwidth]{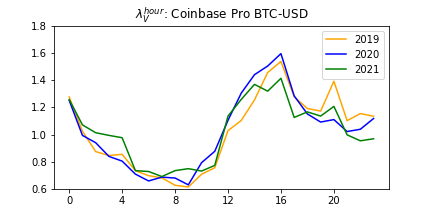}
\par\end{centering}
\begin{centering}
\includegraphics[width=0.5\textwidth]{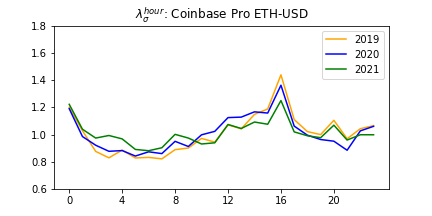}\includegraphics[width=0.5\textwidth]{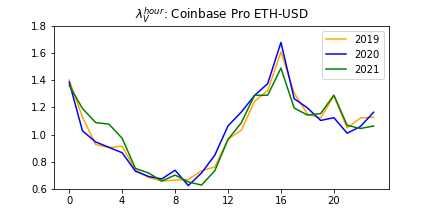}
\par\end{centering}
\centering{}\includegraphics[width=0.5\textwidth]{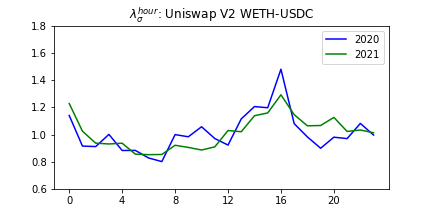}\includegraphics[width=0.5\textwidth]{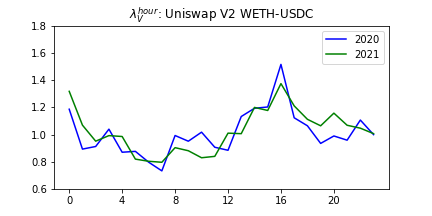}
\begin{small}
\caption{Hour-of-the-Day relative volatility (left panels) and relative volume (right panels) by calendar year. The sample period for Binance and Coinbase Pro is  January 1, 2019 to May 31, 2021 
and the sample period for Uniswap V2 is October 1, 2020 to May 31,
2021.\label{fig:HourOfDaybyYear}}
\end{small}
\end{figure}

\begin{figure}[H]
\begin{centering}
\includegraphics[width=0.5\textwidth]{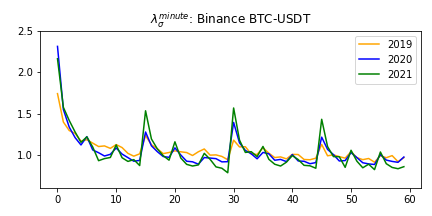}\includegraphics[width=0.5\textwidth]{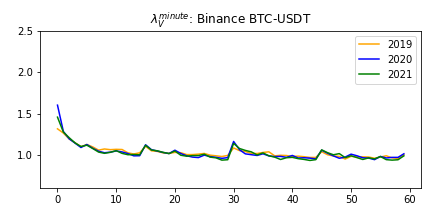}
\par\end{centering}
\begin{centering}
\includegraphics[width=0.5\textwidth]{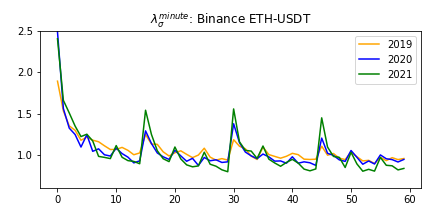}\includegraphics[width=0.5\textwidth]{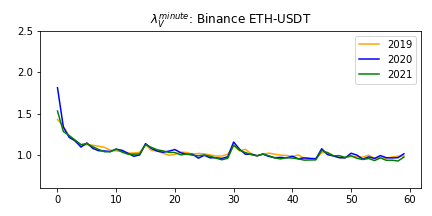}
\par\end{centering}
\begin{centering}
\includegraphics[width=0.5\textwidth]{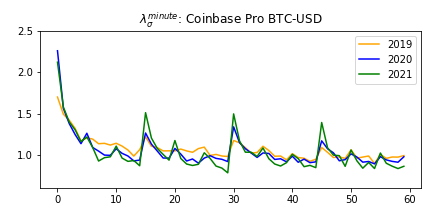}\includegraphics[width=0.5\textwidth]{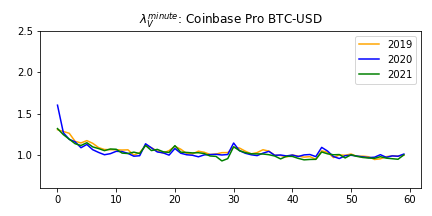}
\par\end{centering}
\begin{centering}
\includegraphics[width=0.5\textwidth]{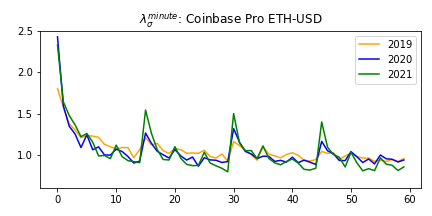}\includegraphics[width=0.5\textwidth]{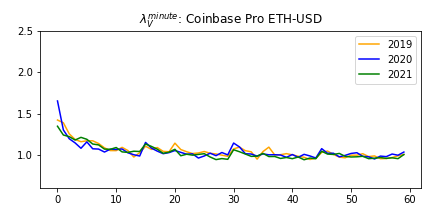}
\par\end{centering}
\centering{}\includegraphics[width=0.5\textwidth]{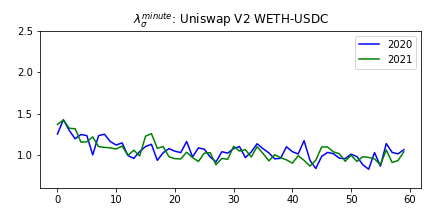}\includegraphics[width=0.5\textwidth]{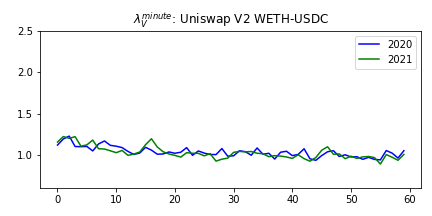}
\begin{small}
\caption{Minute-of-the-Hour relative volatility (left panels) and relative volume (right panels) by calendar year. The sample period for Binance and Coinbase Pro is  January 1, 2019 to May 31, 2021 
and the sample period for Uniswap V2 is October 1, 2020 to May 31,
2021.\label{fig:MinuteOfHourbyYear}}
\end{small}
\end{figure}

\begin{figure}[H]
\begin{centering}
\renewcommand{\tabcolsep}{0pt}
\begin{tabular}{ p{0.02\textwidth}p{0.49\textwidth}p{0.49\textwidth} }
& \multicolumn{1}{c}{Volatility (BTC/USDT, Binance)}&\multicolumn{1}{c}{Volume (BTC/USDT, Binance)}\\
\rotatebox{90}{\makebox[0.32\textwidth]{\Large 2019}}&
\includegraphics[width=0.49\textwidth]{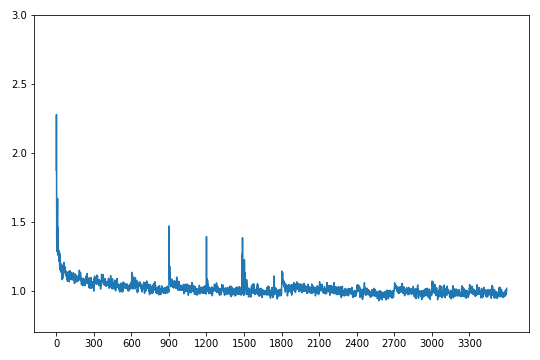}%
\!\!\!%
&
\includegraphics[width=0.49\textwidth]{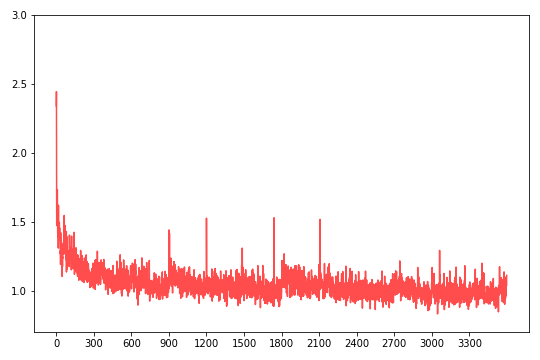}%
\\
\rotatebox{90}{\makebox[0.32\textwidth]{\Large 2020}}&
\includegraphics[width=0.49\textwidth]{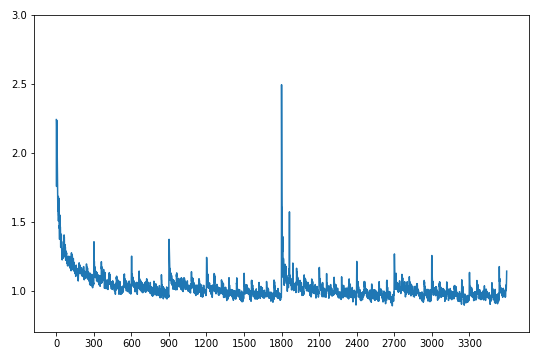}%
\!\!\!%
&
\includegraphics[width=0.49\textwidth]{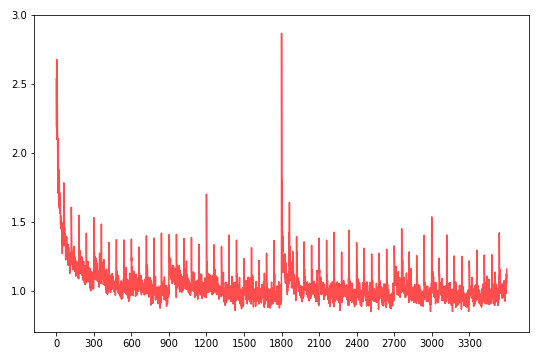}%
\\
\rotatebox{90}{\makebox[0.32\textwidth]{\Large 2021}}&
\includegraphics[width=0.49\textwidth]{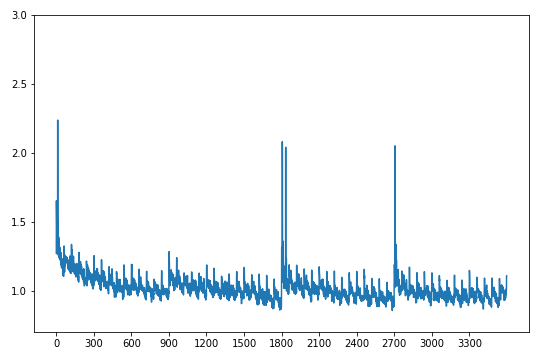}%
\!\!\!%
&
\includegraphics[width=0.49\textwidth]{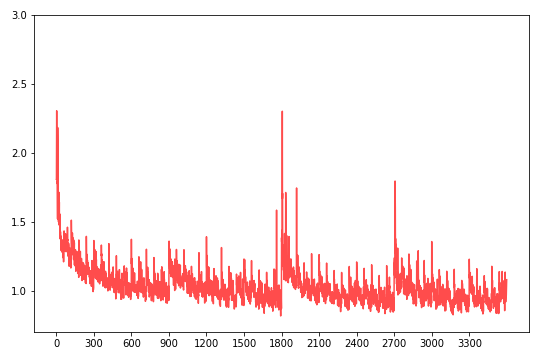}%
\end{tabular}
\renewcommand{\tabcolsep}{6pt}
\par\end{centering}
\begin{small}\caption{Distribution of volatility and volume over the seconds-of-the-hour for BTC/USDT traded on Binance (by calendar year). The left panels show the distribution of volatility and the right panels are the corresponding results for volume. \label{fig:Second-of-Hour-Binance}}\end{small}
\end{figure}

\begin{figure}[H]
\begin{centering}
\renewcommand{\tabcolsep}{0pt}
\includegraphics[width=0.90\textwidth]{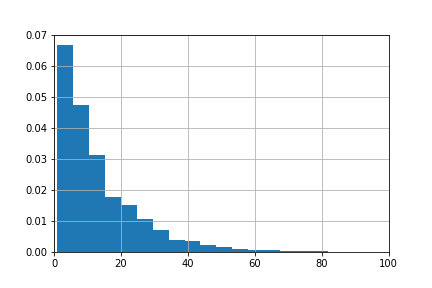}%
\par\end{centering}
\begin{small}\caption{Distribution of the time it takes to produce a new block on the Ethereum blockchain. \label{fig:BlockDurations}}\end{small}
\end{figure}

Figure \ref{fig:BlockDurations} shows the distribution of Ethereum block time over the sample period from October 1, 2020 to May 31, 2021. A block time is a duration, defined by the time it takes to add a new block to the blockchain. The average block time is 13.3 seconds in this sample period.

\begin{figure}[H]
\begin{centering}
\includegraphics[width=0.47\textwidth]{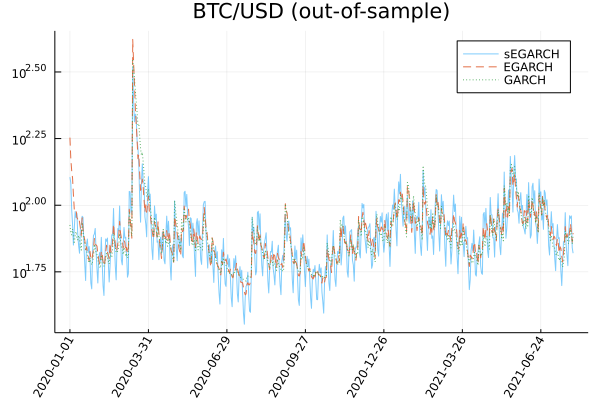}%
\includegraphics[width=0.47\textwidth]{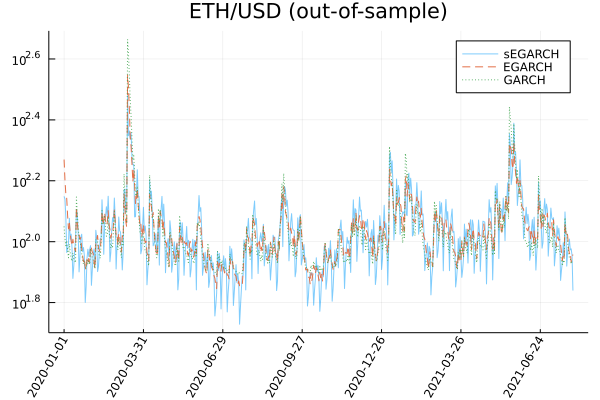}%
\par
\includegraphics[width=0.47\textwidth]{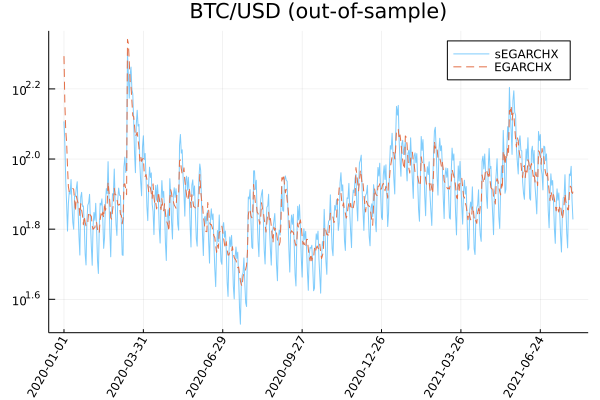}%
\includegraphics[width=0.47\textwidth]{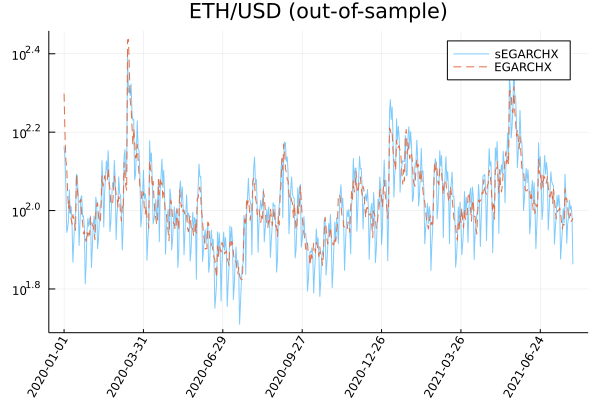}%
\par\end{centering}
\begin{small}\caption{Out-of-sample conditional annualized volatility for each of the five estimated GARCH models. Left panels are for BTC/USD and right panels are for ETH/USD. The upper panels are for the three models that do not utilize the realized variance and the lower panels are for GARCHX models.\label{fig:GARCHmodels}}
\end{small}
\end{figure}

The panels in Figure \ref{fig:GARCHmodels} show the conditional variances in units of annualized volatility $\sqrt{365\cdot \hat{h}_t}$ for the out-of-sample period. The left two panels are the results for BTC/USD and the right two panels are for ETH/USD. The upper panels are the models that do not utilize $\mathrm{RV}_t$ in the modeling and the lower panels are the two variants of the EGARCHX model which include the realized variance in the GARCH equation.

\section{Computation of Confidence Intervals}

In Figures \ref{fig:Day-of-Week-Relative}, \ref{fig:Hour-of-Day-Relative} and \ref{fig:Minute-of-Hour-Relative} we present the relative shares of volatility, $\lambda_{\sigma}$,
for day-of-the-week, hour-of-the-day, and minute-of-the-hour, along
with confidence bands (shaded regions). Similarly, in Figure \ref{fig:Hour-of-Day-Illiquidity} we include confidence bands for the relative illiquidity measure. Here, we detail how the confidence bands are computed and present the justification for the expressions from which these are computed. The relative volatility measures, presented in Figure \ref{fig:Day-of-Week-Relative} can be written,
\(\lambda_{\sigma}^{\mathrm{day}}(d)=7\times \hat\theta_d \) where \[\hat\theta_d = \frac{1}{N_{d}}\sum_{w=1}^{N_d} \tfrac{X_{w,d}}{X_{w,d}+Y_{w,d}}\] 
with \(X_{w,d}=\sum_{h,m}|y_{\tau(w,d,h,m)}|\) and $Y_{w,d} = \sum_{i=1}^{6}\sum_{h,m}|y_{\tau(w,d-i,h,m)}|$

We define $\theta_d = \mathbb{E}Z_{d,w}$ where $Z_{d,w}=\tfrac{X_{w,d}}{X_{w,d}+Y_{w,d}}$ and, provided that $Z_{w,d}$ satisfies the conditions of a central limit theorem, it follows that
\[
\sqrt{W}(\hat{\theta}_{d}-\theta_{d})\overset{d}{\rightarrow}N(0,\omega^{2}).
\]
where the variance, $\omega^2$, can be estimated by 
\[
\hat{\omega}_{d}^{2}=\frac{1}{N_d}\sum_{w=1}^{N_d}\left(Z_{d,w}-\hat{\theta}_{d}\right)^{2}.
\]
The parameter $\theta_d$ as well as the variables $Z_{w,d}$ are bounded between zero and one as is the case for a parameter that represents a probability. For parameters of this type, it is often possible to construct confidence intervals with better finite sample properties by applying the logistic transformation, so define $\vartheta_d = \mathrm{logit}(\theta_d)$ and $\hat\vartheta_d = \mathrm{logit}(\hat\theta_d)$, where $\mathrm{logit}(x)=\log\tfrac{x}{1-x}$. Then, by the delta-method, we have
\[
\sqrt{N_d}(\hat\vartheta_d-\vartheta_d)\overset{d}{\rightarrow}N(0,\sigma^2),\qquad\text{with}\quad\sigma^2=\omega^{2}/[\theta_{d}(1-\theta_{d})]^{2}.
\]
Confidence intervals for $\vartheta_d$ are now obtained with
\(
\hat\vartheta_d \pm 1.96 \hat\sigma_\vartheta\), where \(
\hat\sigma_\vartheta = \tfrac{1}{\sqrt{N_d}}\hat\omega/\hat\theta_d(1-\hat\theta_d)\). These are translated to to confidence intervals for $\theta_d$ by applying the inverse mapping, $\mathrm{tigol}(y)=1/(1+e^{-y})$, to the endpoints of the confidence interval for $\vartheta_d$. Finally, the confidence interval for
$\lambda_{\sigma}^{\mathrm{day}}(d)$ is now given by
\[
\left[7\times\mathrm{tigol}(\hat\vartheta_d - 1.96 \hat\sigma_\vartheta),7\times\mathrm{tigol}(\hat\vartheta_d + 1.96 \hat\sigma_\vartheta)\right].
\]
The confidence intervals for $\lambda_{\sigma}^{\mathrm{hour}}(h)$,  $\lambda_{\sigma}^{\mathrm{minute}}(m)$, and $\lambda_{\mathrm{Illiquid}}^{\mathrm{hour}}(h)$ are constructed similarly. 

\end{document}